\documentclass[a4paper, 11pt]{article}
\usepackage[utf8]{inputenc}
\usepackage[T1]{fontenc}
\usepackage{lmodern}
\usepackage{microtype}
\usepackage[left=2.5cm, right=2cm, top=2.5cm, bottom=2cm]{geometry}

\usepackage{amsmath}
\usepackage{hyperref}
\usepackage{graphicx, bm, color}

\usepackage{wasysym}
\usepackage{verbatim}
\usepackage{amssymb}
\usepackage{braket}
\usepackage{epstopdf}
\usepackage{animate}
\usepackage{xmpmulti}
\usepackage{centernot}
\usepackage{multirow}
\usepackage{tikz-cd}
\usetikzlibrary{calc}
\usetikzlibrary{intersections}
\usetikzlibrary{shapes.geometric}
\usetikzlibrary{positioning}
\usepackage[graphicx]{realboxes}
\usepackage{float}
\usepackage{mathtools}
\usepackage{comment}
\usepackage{braket, tensor}
\usepackage{cite}
\usepackage{adjustbox}
\usepackage[normalem]{ulem}

\def\co{\Delta}
\newcommand{\be}{\begin{equation}}

\newcommand{\ee}{\end{equation}}
\newcommand{\beq}{\begin{eqnarray}}
\newcommand{\eeq}{\end{eqnarray}}
\newcommand{\ba}{\begin{align}}
\newcommand{\ea}{\end{align}}

\newcommand{\up}{\uparrow}
\newcommand{\down}{\downarrow}

\begin{document}
    \begin{center}
        \baselineskip 24 pt {\LARGE \bf Quantum deformations of $\mathcal{U}(\mathfrak{sl}(2, \mathbb{R}))$. Part I: Fidelity and experimental benchmarking}
    \end{center}

    \begin{center}
        {V. Mariscal$^{1}$, J.J. Relancio$^{1,2}$ and L. Santamar\'ia-Sanz$^{1}$}

        \medskip

        {$^{1}$Departamento de Matem\'aticas y Computaci\'on, Universidad de Burgos, 09001 Burgos, Spain}

        {$^{2}$Centro de Astropart\'{\i}culas y F\'{\i}sica de Altas Energ\'{\i}as (CAPA), Universidad de Zaragoza, Zaragoza 50009, Spain}
        \medskip

        e-mail:
        {\href{mailto:vmariscal@ubu.es}{vmariscal@ubu.es}, \href{mailto:jjrelancio@ubu.es}{jjrelancio@ubu.es}, \href{mailto:lssanz@ubu.es}{lssanz@ubu.es}}
    \end{center}

    \begin{abstract}
        This work explores the effects of both the standard quantum $q$-deformation and the non-standard
        $h$-deformation of the Hopf algebra $\mathcal{U}(\mathfrak{sl}(2, \mathbb{R}))$ on multi-qubit systems. By constructing the states of a Hilbert space of $N$ qubits through the Clebsch–Gordan coefficients associated with the deformed algebras, we show that these states naturally coincide with the eigenstates of the Hamiltonian of the $q$- and $h$-deformed Kittel–Shore models.  We compare the resulting deformed states with those typically targeted in quantum information experiments, providing a bridge between algebraic constructions and experimentally relevant quantum resources. Fidelities with respect to the undeformed states are computed to establish how the quantum correlations are affected, both for few-qubit systems (including Dicke and non-Dicke states), and in the macroscopic limit ($N \to \infty$) through closed-form formulas derived for arbitrary Dicke states. The results reveal different behaviors between the two
        deformations. The $q$-deformation smoothly modifies the states and maintains a residual overlap with the original configurations, while the $h$-deformation rapidly makes  the states orthogonal to their undeformed counterparts. Both models demand a standard $N^{-1}$ rescaling to preserve fidelity stability in the macroscopic limit.
    \end{abstract}
    \thispagestyle{empty}


    \section{Introduction}

    Quantum groups \cite{Drinfeld,Jimbo:1985zk} represent a generalization of classical
    symmetry groups in quantum systems and provide a powerful mathematical
    framework for describing  quantum phenomena. Their duals are algebraic structures
    emerging from the deformation of Lie algebras, and appear in various areas of theoretical
    physics and mathematics \cite{gawedzki1991classical,isaev1994quantum}. In
    quantum systems, quantum groups offer a method to describe symmetries that are
    not captured by traditional group theory. They incorporate noncommutative
    geometry and quantum deformations, allowing the exploration of quantum systems
    with complex symmetry properties.

    Among the deformations of symmetry algebras, the most widely studied is the
    standard or $q$-deformation, introduced in the context of quantum groups by Drinfel’d
    and Jimbo in the 1980s \cite{drinfel1990hopf,Jimbo:1985zk}. This construction
    relies on the introduction of a parameter $q$, which modifies both the commutation relations and the Hopf algebra structure,  and can be formulated for a broad class of Lie algebras. In this work, we focus on the case of
    $\mathcal{U}(\mathfrak{sl}(2, \mathbb{R}))$, leading to the quantum algebra $\mathcal{U}_{q}(
    \mathfrak{sl}(2, \mathbb{R}))$. The $q$-deformation has a rich
    interpretation in representation theory and has found applications in
    integrable systems and spin models in statistical physics
    \cite{kassel2012quantum,Chari,kulish1993quantum,kulish1991general}. In the limit
    $q\rightarrow 1$, the undeformed algebra is recovered. For $q\neq 1$,
    however, the Clebsch--Gordan coefficients are replaced by their $q$-analogs,
    which are expressed in terms of basic hypergeometric series and $q$-factorials.
    Consequently, the construction of collective states is affected \cite{biedenharn1989quantum,kirillov1990uq}. In addition to this standard deformation, there also exists a non-standard
    or $h$-deformation, known as the Jordanian deformation \cite{Majid:1988we,KulishReshetikhin}. This construction relies on the introduction of a parameter $h$, which modifies the Hopf algebra structure but not the
commutation relations. In the limit $h \rightarrow 0$ the undeformed    algebra is also recovered. Unlike the standard $q$-deformation, which is governed by a quasitriangular $R$-matrix,
    the parameter $h$ in this non-standard case is introduced through a purely
    triangular $R$-matrix. This structure corresponds to a nilpotent solution of
    the classical Yang--Baxter equation, typically generated via a Drinfeld twist
    \cite{AngelBallesteros_1996}. Consequently, the structure of $\mathcal{U}_{h}
    (\mathfrak{sl}(2, \mathbb{R}))$ exhibits coproducts that
    are essentially different from those of the $q$-deformation. 
    

    A comparative study between standard and non-standard
    deformations is of great interest, as both provide alternative algebraic frameworks
    to explore how the symmetry of a system controls the properties of quantum
    states. In particular, while $q$-deformations are well established in the theory
    of quantum groups and their physical applications, $h$-deformations have
    shown great potential as both a mathematical and physical laboratory for
    generating new classes of quantum states \cite{ballesteros2025entangled},
    with tunable properties through the deformation parameter $h$. Focusing 
our analysis on the fundamental spin-1/2 representation, both deformations allow for the systematic generation of new classes of multi-qubit quantum states. The algebraic construction of the $q$-deformed and $h$-deformed states finds its natural physical realization as the energy eigenstates of the deformed  Kittel--Shore (KS) model. The undeformed Kittel--Shore Hamiltonian is given by \cite{kittel1965development}
    \begin{equation}
        \hat{H}_{KS}=- I \sum_{i<j}^{N}\Vec{J}_{i}\cdot\Vec{J}_{j}, \label{eq:hamiltonian1}
    \end{equation}
    where $I$ is the interaction constant between the spins and $\Vec{J}_{i}$ represents the total angular momentum operator of the $i$-th site. The KS model describes a system of $N$ qubits characterized by long-range all-to-all interactions where every site is coupled identically to all others. This globally connected  architecture is governed by an underlying $\mathcal{U}(\mathfrak{sl}(2, \mathbb{R}))$ coalgebra
    symmetry, which dictates the dynamical evolution of the system. The $q$-KS model has been described in \cite{Ballesteros:2025cia}. Here, we present the $h$-KS model for the first time and discuss both deformations, including the differences between their corresponding quantum states.

    Due to its ability to describe
    systems where spins exhibit isotropic exchange owing to constant long-range
    coupling, the undeformed KS model is particularly
    suited to characterize small clusters of equidistant spins (dimers,
    equilateral triangles, regular tetrahedra), allowing the study of the magnetic properties of
    organic molecules containing embedded paramagnetic ions
    \cite{ciftja1999equation}, the melting behavior of small clusters
    in porous media \cite{sheng1981melting}, as well as the collective properties
    emerging from individual magnetic moments in ultra-small spin systems \cite{ciftja2001irregular, ciftja2000spin}. In the field of condensed matter and material science, 
    the KS model is used to model high-temperature superconductors \cite{al2004nonlinear}, to study magnetization in layers where spins are identically coupled \cite{al2004extended, gros1995transition, czachor2001green}, and to thermodynamically analyze multiferroic thermomagnetic generators \cite{sandoval2015thermodynamic} (combining the efficiency model described in \cite{hsu2011thermomagnetic} with the  model of Curie point suppression of \cite{sun2004coordination}) to characterize transition behaviors in nickel-based nanosolids. These applications show how the model connects microscopic spin dynamics and
    macroscopic thermodynamic efficiency. Moreover, the KS model is also currently used in quantum information processing and the study of non-equilibrium phases of matter. Its structure provides a natural description for qubits implemented in
    ammonia-based molecular gates \cite{ferguson2002ammonia} or electronic spins
    in semiconductor quantum dots \cite{woodworth2006few}, where few-body interactions
    must be carefully tuned to circumvent decoherence and multi-body effects \cite{benenti2004principles}. It is also a candidate for the realization of discrete-time crystals in superconducting circuits and cold atom traps \cite{barfknecht2019realizing}. In this paper, we compare the experimental states with those theoretically obtained from the deformed models, showing that both of them can resemble some features of the errors and noise of the state constructions.  
    

    Quantum fidelity serves as a fundamental statistical measure used to quantify the similarity between two quantum states, effectively functioning as a metric for information preservation in quantum communication and computation \cite{nielsen_chuang_2010}. For pure states, represented by state vectors $|\psi\rangle$ and $|\phi\rangle$, the fidelity is defined as the squared modulus of their inner product, $F(|\psi\rangle, |\phi\rangle) = |\langle\psi|\phi\rangle|^2$, which ranges from unity for identical states to zero for those that are perfectly orthogonal. In the more general context of mixed states described by density operators $\rho$ and $\sigma$, the fidelity is generalized \cite{jozsa1994} through $F(\rho, \sigma) = (\text{Tr}\sqrt{\sqrt{\rho}\sigma\sqrt{\rho}})^2$, providing a geometric interpretation of the distance between states within the Hilbert space while accounting for statistical uncertainty. This measure is essential for benchmarking the accuracy of quantum gates and state preparation, as it characterizes the deviation of an experimental result from its theoretical target due to environmental decoherence or operational errors.  In this work, we study the fidelity to compare the results for some
    undeformed and deformed states, obtained from deformed $q$- and $h$-KS Hamiltonians. Particular attention will be focused on Dicke states~\cite{dicke1954coherence}, which are a prominent class of quantum states characterized as the fully symmetric eigenstates of total spin operators in SU(2) models. Their primary mathematical property is their symmetry, as they remain unchanged (invariant) under the operation of the permutation group.  For a system of $N$ qubits with $k$
    excitations, the corresponding Dicke state $\ket{D_N^{(k)}}$ in the undeformed theory
    reads
    \begin{equation}
             \vert D^{(k)}_{N}\rangle = \frac{1}{\sqrt{\binom{N}{k}}}\sum_{S \in \mathcal{B}_{N,k}}
        |S\rangle, \label{eq:dicke_def}
    \end{equation}
     where $\mathcal{B}_{N,k}$ represents the set of all binary words of length $N$ and Hamming
    weight $k$ (that is, all possible states with exactly $k$ spins $\up$ and $N-k$ spins $\down$)
    \cite{nepomechie2023qudit}. In quantum computing, Dicke states are highly prized~\cite{bartschi2019deterministic,mukherjee2020actual,aktar2022divide} due to their symmetry, which allows them to simultaneously offer strong multipartite entanglement and protection against particle deletion. Furthermore, they are essential for achieving metrological enhancement in sensitive measurements~\cite{Wieczorek2009,Prevedel2009,bengtsson2017geometry,walter2016multipartite}.We focus our attention on Dicke states, yet the analysis is not restricted to them. The quantum algebraic approach used in this work also identifies states within other representation, labeled here as $M$, $V$, $T$, $R$, and $Q$ representations. These states are of significant interest as they manifest intrinsically different algebraic symmetries.  The fundamental distinction between the $q$ and $h$ deformations manifests in their asymptotic structural impact on the state space. On one hand, the $q$-deformation operates as a smooth algebraic perturbation, ensuring that the deformed states retain a non-vanishing residual fidelity for large values of $q$. On the other hand, the Jordanian $h$-deformation triggers a severe, state-dependent orthogonalization governed entirely by the internal combinatorics of the spin lattice. Remarkably, for $q$- and $h$-deformed Dicke states of $N$ qubits, the fidelity per site in the thermodynamic limit is computed, showing that a $N^{-1}$ rescaling in the deformation parameter is needed to avoid orthogonality catastrophe. This rescaling was also obtained in \cite{Ballesteros:2025cia} in order to assure spectral and thermodynamic stability of the $q$-KS model.

    The remainder of this paper is organized as follows. Sec.~\ref{sec:revisiting}
    delves into the mathematical structure of both undeformed and deformed
    $\mathcal{U}_{q}(\mathfrak{sl}(2, \mathbb{R}))$ and
    $\mathcal{U}_{h}(\mathfrak{sl}(2, \mathbb{R}))$ algebras, examining their commutation
    relations, coproducts, and representations. It also addresses the
    construction of Clebsch--Gordan coefficients to develop multi-particle deformed quantum states.
    In Sec.~\ref{sec:kittel_shore}, we introduce the Kittel--Shore model and construct
    the corresponding superintegrable deformations. In Sec.~\ref{sec:comparison},
    we compare these theoretical predictions with recent experimental results on
    quantum state preparation. Sec.~\ref{sec:entropy} offers a comparative
    analysis of the fidelity for undeformed, $q$-deformed, and $h$-deformed
    quantum states of qubits. Finally, we conclude with some
    insights and future directions in Sec.~\ref{sec:conclusions}.

    \section{Revisiting
    \texorpdfstring{$\mathcal{U}(\mathfrak{sl}(2, \mathbb{R}))$, $\mathcal{U}_{q}
    (\mathfrak{sl}(2, \mathbb{R}))$ and $\mathcal{U}_{h}(\mathfrak{sl}(2, \mathbb{R}
    ))$}{U(sl(2,R)), Uq(sl(2,R)) and Uh(sl(2,R))}
    Hopf algebras}
    \label{sec:revisiting}

    The Lie algebra $\mathfrak{sl}(2, \mathbb{R})$ is a fundamental component in
    the description of quantum systems with rotational symmetry, such as spins.
    Its algebraic properties are defined by the following commutation relations of the generators $\{J_{z}, J_+, J_-\}$:
    \begin{equation}
        [J_{z},J_{\pm}]= \pm J_{\pm},\qquad [J_{+},J_{-}]= 2 J_{z}, \label{eq:su2Lie}
    \end{equation}
    where
    \begin{equation}
        J_{x}= \frac{1}{2}\sigma_{x}, \qquad J_{y}= \frac{1}{2}\sigma_{y}, \qquad
        J_{z}= \frac{1}{2}\sigma_{z}, \qquad J_{\pm}= J_{x}\pm i J_{y},
    \end{equation}
    gives the fundamental $2 \times 2$ finite-dimensional irreducible representation
    in terms of the Pauli matrices.

    To describe an $N$-particle system, the single-spin algebraic structure must be extended through a formal composition rule that preserves the underlying symmetry across the many-body Hilbert space.  This framework is provided by the Hopf algebraic structure of the universal enveloping algebra $\mathcal{U}(\mathfrak{sl}(2, \mathbb{R}))$, which defines a consistent mapping for collective operators. Beyond the standard Lie algebra axioms, the Hopf structure includes a coproduct ($\Delta$),
    counit ($\epsilon$), and antipode ($S$). These operations are indispensable for consistently defining and manipulating tensor products across representation spaces. Coproduct, in particular,
    describes how the algebra's generators act on a composite state. For
    $\mathcal{U}(\mathfrak{sl}(2, \mathbb{R}))$, it is defined as:
    \begin{equation}
        \Delta (J_{i})=J_{i}\otimes 1+1\otimes J_{i}.
    \end{equation}
    This relationship allows the calculation of the collective properties of a
    particle system from the individual properties of its components.

    The decomposition of a composite state into irreducible representations is
    critical in quantum mechanics \cite{Sakuraibook, cohenbook}. This task is accomplished using Clebsch--Gordan
    coefficients, which act as the expansion coefficients in a linear combination. They act as scalars in a linear combination that transforms the tensor product basis of two (or more) irreducible representations into a new basis that decomposes the space into a direct sum of irreducible representations:
    \begin{eqnarray}
        \label{boots1} \ket{j,m} =\sum_{m=m_1+m_2} \mathcal{C}^{j_1,j_2,j}_{m_1,m_2,m} \ket{j_1,m_1}\otimes
        \ket{j_2, m_2},
    \end{eqnarray}
    where $j,j_1,j_2$ are non-negative integers or half-integers such that $j=(j_{1}+ j_{2})\,\dots,|j_{1}-j_{2}|$, and $m_{i}=-j,...,j$.
    $\mathcal{C}^{j_1,j_2,j}_{m_1,m_2,m}$ are the associated Clebsch--Gordan coefficients, which can be written as

    \begin{eqnarray}
        &&\label{boots2} \mathcal{C}^{j_1,j_2,j}_{m_1,m_2,m}=\delta_{m_1+m_2,m} \left(\frac{
        (j_{1}+j_{2}-j)! (j_{1}-j_{2}+j)! (-j_{1}+j_{2}+j)! }{(j_{1}+j_{2}+j+1)!}
        \right)^{1/2} \nonumber\\
        &&\times \left((j_1+m_1)!(j_1-m_1)!(j_2+m_2)!(j_2-m_2)!(j+m)!(j-m)!(2j+1)
        \right)^{1/2} \nonumber\\
        &&\times  \sum_k
        \frac{(-1)^{k}}{k!(j_{1}+j_{2}-j-k)!(j_{1}-m_{1}-k)!(j_{2}+m_{2}-k)!(j-j_{2}+m_{1}+k)!
        (j-j_{1}-m_{2}+k)!},\nonumber\\
    \end{eqnarray}
with $\delta_{a,b}$ the Kronecker delta.

    \subsection{The standard deformation}

    The Hopf algebra
    $\mathcal{U}_{q}(\mathfrak{sl}(2, \mathbb{R}))$ is generated by
    $\{L_{z},L_{+},L_{-}\}$ and can be defined in terms of the following
    relations~\cite{Biedenharnbook} where $q\in \mathbb{C} \setminus \{ e^{i \frac{2\pi \, r}{n}} : n \in \mathbb{N}, r = 0, 1, \dots, n-1 \}$:
    \begin{equation}
        [L_{z},L_{\pm}]= \pm L_{\pm},\qquad [L_{+},L_{-}]= \left[ 2 L_{z}\right]_{q}
        = \frac{q^{L_z}-q^{-L_z}}{q^{1/2}-q^{-1/2}}. \label{suq2}
    \end{equation}
    Above, we have made use of the symbol for $q$-numbers
    \begin{equation}
        [\ell]_{q}\coloneqq \frac{q^{\ell/2}-q^{-\ell/2}}{q^{1/2}-q^{-1/2}}=\frac{\sinh(\eta
        \ell/2)}{\sinh(\eta/2)}, \quad \ell\in\mathbb{Z} \label{eq:qnumber},
    \end{equation}
where the usual convention\cite{Jimbo:1985zk} $q=e^\eta$ has been used. 
    The coalgebra structure for the
    $\mathcal{U}_{q}(\mathfrak{sl}(2, \mathbb{R}))$ algebra is generated by the
    deformed coproduct map
    \begin{equation}
        \Delta_{q}(L_{\pm})= q^{-L_z/2}\otimes L_{\pm}+ L_{\pm}\otimes q^{L_z/2},
        \qquad \Delta_{q}(L_{z})= 1 \otimes L_{z}+ L_{z}\otimes 1. \label{qcop}
    \end{equation}
    Irreducible components in the tensor product (given by the deformed coproduct
    map~\eqref{qcop}) of two $\mathcal{U}_{q}(\mathfrak{sl}(2, \mathbb{R}))$ representations
    are again obtained in terms of \eqref{boots1}, where the $q$-Clebsch--Gordan
    coefficients (see~\cite{alvarez2024russian} and references therein) are used:
    \begin{equation}
        \begin{split}
            {\cal C}_{m_1, m_2, m}^{j_1, j_2, j}(q)\,=&\,\delta_{m,m_1+m_2}q^{\frac{1}{2}
            (j_1 m_2-j_2 m_1)-\frac{1}{4} (-j+j_1+j_2) (j+j_1+j_2+1)}\\
            &\sqrt{\frac{[2 j+1]_{q}[j+m]_{q}! [j_{2}-m_{2}]_{q}![j+j_{1}-j_{2}]_{q}!
            [-j+j_{1}+j_{2}]_{q}! [j+j_{1}+j_{2}+1]_{q}! }{[j-m]_{q}! [j_{1}-m_{1}]_{q}!
            [j_{1}+m_{1}]_{q}! [j_{2}+m_{2}]_{q}! [j-j_{1}+j_{2}]_{q}! }}\\
            &\sum_{n=0}^{\min (-j+j_1+j_2,j_2-m_2)}\frac{ [2 j_{2}-n]_{q}! (-1)^{-j+j_1+j_2+n}q^{\frac{1}{2}
            n (j_1+m_1)}[j_{1}+j_{2}-m-n]_{q}! }{ [n]_{q}! [j_{2}-m_{2}-n]_{q}! [-j+j_{1}+j_{2}-n]_{q}!
            [j+j_{1}+j_{2}-n+1]_{q}! }\, .
        \end{split}
        \label{eq:qcg}
    \end{equation}
  Using these coefficients, the quantum states are obtained. While cases for up to 3 qubits can be found in \cite{li2015entanglement,Ballesteros:2025cia}, the newly derived 4-qubit states, along with the previous ones, are collected in Appendix~\ref{app1:sec1}.

    \subsection{The non-standard deformation}
    The Hopf algebra $\mathcal{U}_h(\mathfrak{sl}(2, \mathbb{R}))$ is generated by $\{H, Z_+, Z_-\}$ together with the usual commutators for $\mathfrak{sl}
    (2, \mathbb{R})$ of Eq. \eqref{eq:su2Lie}, providing $H=2J_{z}, \,
    Z_{\pm}= J_{\pm}$. The generators $\{H, Z_{\pm}\}$ act on the computational
    basis exactly the same as the generators of
    $\mathcal{U}(\mathfrak{sl}(2, \mathbb{R}))$ \cite{JVanderJeugt1998}.
    Therefore, the entire effect of the deformation is encoded by the action on multipartite
    states.

    The $\mathcal{U}_{h}(\mathfrak{sl}(2, \mathbb{R}))$ co-algebra structure is given
    by the coproduct map, which reads \cite{Aizawa1997}
    \begin{eqnarray}
        \Delta_h(H) &= & H\otimes 1 + 1 \otimes H + 2H \otimes \sum_{n=1}^\infty
        \left(\frac{hZ_{+}}{ 2} \right)^n + \sum_{n=1}^\infty \left(- \frac{hZ_{+}}{
        2} \right)^n \otimes 2H,\nonumber\\
        \Delta_h(Z_+) &= & (1\otimes Z_+ + Z_+ \otimes 1)\left(\sum_{n=0}^\infty
        \left(-\frac{h^{2}}{4}\right)^n \; Z_+^n \otimes Z_+^n \right),\nonumber\\
        \Delta_h(Z_-) &= & Z_- \otimes \sum_{n=0}^{\infty} (n+1) \left(\frac{h Z_{+}}{2}
        \right)^n + \sum_{n=0}^{\infty}\; (n+1) \left(-\frac{h Z_{+}}{2} \right)^n
        \otimes Z_- \nonumber \\
        & &+ h\left(C_h-{\frac{H^{2}}{ 4}} \right) \otimes \sum_{n=1}^{\infty} n
        \left(\frac{h Z_{+}}{2} \right)^n - \sum_{n=1}^{\infty} n \left( -\frac{h
        Z_{+}}{2} \right)^n \otimes h \left(C_h-{\frac{H^{2}}{ 4}}\right) \nonumber
        \\
        & &+ \left({\frac{h}{ 2}}\right)^2 Z_+Z_-Z_+ \otimes \sum_{n=2}^{\infty}
        (n-1) \left(\frac{h Z_{+}}{2} \right)^n + \sum_{n=2}^{\infty} (n-1)
        \left(-\frac{h Z_{+}}{2} \right)^n \otimes \left({\frac{h}{ 2} }\right)^2
        Z_+Z_-Z_+,\nonumber\\
        \label{eq:coproduct_j}
    \end{eqnarray}
    where $C_{h}= Z_{+}Z_{-}+ H^{2}/4 -H/2$ is the Casimir element. In the limit
    $h \to 0$ we recover the undeformed coproduct.

    The Clebsch--Gordan coefficients for
    $\mathcal{U}_{h}(\mathfrak{sl}(2, \mathbb{R}))$ (denoted by ${\cal C}^{j_1,j_2,j}
    _{m_1,m_2,m}(h)$) can be derived from those of
    $\mathcal{U}(\mathfrak{sl}(2, \mathbb{R}))$ by means of
    \cite{JVanderJeugt1998}
    \begin{eqnarray}
        \label{program1} &&{\cal C}^{j_1,j_2,j}_{m_1,m_2,m}(h) = \sum_{m_1+m_2=m}
        \mathcal{C}^{j_1,j_2,j}_{m_1,m_2,m} A^{m_1,m_2}_{n_1-m_1,n_2-m_2},
    \end{eqnarray}
    where
    \begin{eqnarray}
        \label{program2} && A^{m_1,m_2}_{r,l} = a^{m_1,m_2}_{r,l}
        \frac{\alpha_{j_1,m_1+r}\, \alpha_{j_2,m_2+l}}{ \alpha_{j_1,m_1}\alpha_{j_2,m_2}},\nonumber\\
        && \alpha_{j,m}= \sqrt{\frac{(j+m)!}{(j-m)!}},\nonumber \\
        && a_{r,l}^{m_1,m_2}= (-1)^r 2^{-r-l} h^{r+l} (b_{r,l}^{m_1,m_2}-b_{r-1,l-1}^{m_1,m_2}),
        \\
        &&b^{m_1,m_2}_{r,l} = \left\{
        \begin{array}{ll}
            \frac{(-2m_{1}-r)_{l}(-2m_{2}-l)_{r}}{ r!\,l!} & \hbox{if }r\geq 0\hbox{ and } l\geq 0 , \nonumber \\[2mm]
            0                                            & \hbox{otherwise},
        \end{array}
        \right.
    \end{eqnarray}
    with $(a)_b$ the Pochhammer symbol. Note that the above $h$-Clebsch--Gordan coefficients generally yield unnormalized
    states. In light of the results in \cite{ballesteros2025entangled}, we provide the explicit form of the quantum states in Appendix~\ref{app1:sec2}.

    \section{Kittel--Shore model}
    \label{sec:kittel_shore} The algebraic construction of the aforementioned
    $q$-deformed and $h$-deformed states (see Appendices~\ref{app1:sec1} and~\ref{app1:sec2})
    finds its natural physical realization as the energy eigenstates of the
    deformed KS Hamiltonians.

    A fundamental feature of this model is its maximal superintegrability. In
    the context of many-body physics, a system with $N$ degrees of freedom is considered
    maximally superintegrable if it possesses $2N-1$ independent integrals of motion
    (conserved quantities) that commute with the Hamiltonian, forming a non-abelian
    symmetry algebra \cite{magyari1987integrable,miller2013classical}. Unlike
    many-body systems that require the Bethe-ansatz \cite{Faddeev:1996iy} for their analytical
     resolution, the high degree of symmetry in the Kittel--Shore model allows
    for the energy spectrum to be determined exactly through purely algebraic
    methods. 

    The theoretical justification for extending this framework to the quantum-deformed
    regime rests on the fundamental results established in \cite{ballesteros1998systematic,ballesteros1999integrable,ballesteros2009super}.
    These works prove that the integrability properties of a Hamiltonian system
    are rigorously preserved under a deformation process, provided that the underlying
    coalgebra symmetry remains intact. Since the $q$- and $h$- states are
    derived specifically from the deformed coproduct structures of
    $\mathcal{U}_{q}(\mathfrak{sl}(2, \mathbb{R}))$ and
    $\mathcal{U}_{h}(\mathfrak{sl}(2, \mathbb{R}))$, it is guaranteed that the resulting
    physical system retains its complete set of conserved quantities, ensuring
    that the deformation alters the correlation structure without breaking the
    solvable nature of the model.

    To construct the deformed Kittel--Shore model, the Hamiltonian \eqref{eq:hamiltonian1}
    is expressed in terms of the Casimir operators of the underlying algebra.
    Since the Casimir operator commutes with all generators of the
    algebra—thereby belonging to its center—it serves as the fundamental
    building block for a consistent deformation that preserves the integrability
    of the system. The Kittel--Shore Hamiltonian is thus reformulated as
    \begin{equation}
        \hat{H}_{KS}= -\frac{I}{2}\left(\Delta^{(N)}(C)-\sum_{i=1}^{N}C^{(i)}\right),
    \end{equation}
    where $\Delta^{(N)}$ denotes the $N$-th order coproduct:
       \begin{equation}
\co^{(N)}=(1^{\,\otimes (N-2)}\otimes
\co^{(2)})\circ\co^{(N-1)}= (\co^{(2)}\otimes 1^{\,\otimes (N-2)})\circ\co^{(N-1)}.
\label{fl}
\end{equation}
    By replacing the undeformed
    Casimir operator with its $q$- or $h$-deformed counterpart, we derive the corresponding
    deformed versions of the model. It is worth noticing that the coproduct of the Casimir element is the Casimir of the coproduct. 

    \subsection{The Standard \texorpdfstring{$q$}{q}-Deformation}

    The Casimir operator of $\mathcal{U}
    _{q}(\mathfrak{sl}(2, \mathbb{R}))$  is given by:
    \begin{equation}
        C_{q}= L_{-}L_{+}+ [L_{z}]_{q}[L_{z}+\mathbf{I}]_{q}=L_{+}L_{-}+ [L_{z}]_{q}
        [L_{z}-\mathbf{I}]_{q}\, , \label{qcas}
    \end{equation}
    For a system of $N$ qubits, each local site carries the spin-$1/2$ representation,
    resulting in a local Casimir eigenvalue of $[1/2]_{q}[3/2]_{q}$. Consequently,
    the additive constant in the Hamiltonian effectively scales with the number
    of particles and the deformation parameter ($\sum_{i=1}^{N}C_{q}^{(i)}=N[1/2
    ]_{q}[3/2]_{q}$).

In the absence of an external magnetic field, the energy spectrum of the
undeformed KS model is given by
\begin{equation}
    E(j_{\alpha_i})=-\frac{I}{2}\left(
    j_{\alpha_i}(j_{\alpha_i}+1)-\frac{3N}{4}
    \right),
    \quad i=1,\ldots,r ,
    \label{eigenvaluesKS}
\end{equation}
where \(j_{\alpha_i}\) denotes the total spin of the \(i\)-th irreducible
component appearing in the decomposition of the \(N\)-fold tensor product of
spin-\(1/2\) representations,
\[
\left(D^{1/2}\right)^{\otimes N}
=
D^{j_{\alpha_1}}
\oplus
D^{j_{\alpha_2}}
\oplus
\cdots
\oplus
D^{j_{\alpha_{r}}}.
\]
The label \(\alpha_i\) distinguishes the different irreducible components,
including possible multiple copies with the same value of the total spin.
For \(N\) spin-\(1/2\) representations, the total number of such components is
\[
r=\binom{N}{\lfloor N/2\rfloor}.
\]
Thus, different labels \(\alpha_i\) and \(\alpha_j\) may correspond to the same
spin value, \(j_{\alpha_i}=j_{\alpha_j}\), reflecting the multiplicity of that
irreducible representation. Since the KS Hamiltonian depends only on the total
spin Casimir, the energy is determined by the value of \(j_{\alpha_i}\).

On the other hand, the energy spectrum of the \(q\)-KS model is determined by
\begin{equation}
    E_q(j_{\alpha_i})=-\frac{I}{2}\left(
    [j_{\alpha_i}]_{q}\,[j_{\alpha_i}+1]_{q}
    -N\left[\frac{1}{2}\right]_{q}
    \left[\frac{3}{2}\right]_{q}
    \right),
    \quad i=1,\ldots,r .
    \label{eigenvaluesqKS}
\end{equation}

Concerning the associated eigenvectors, those of the $q$-deformed KS Hamiltonian remain orthogonal, as in the non-deformed case, but their structure is modified by the deformation, which reweights the components in the computational basis through the parameter $q$. This is reflected in the explicit form of the Clebsch--Gordan coefficients of $\mathcal{U}_{q}(\mathfrak{sl}(2, \mathbb{R}))$ in Eq.~(\ref{eq:qcg}) and in the states given in Appendix~\ref{app1:sec1}. At the spectral level, the deformation alters the energy eigenvalues and modifies the probability amplitudes of the computational basis states, thereby reshaping the entanglement structure across the full Hilbert space.

    \subsection{The Jordanian \texorpdfstring{$h$}{h}-Deformation}
    
    The Jordanian Casimir operator $C_{h}$ is the central element of the algebra
    $\mathcal{U}_{h}(\mathfrak{sl}(2, \mathbb{R}))$. In the basis
    $\{Z_{+}, Z_{-}, H\}$, which is related to the classical generators ($\{J_{+}
    ,J_{-},J_{z}\}$) through a non-linear map \cite{Aizawa1997,ballesteros2025entangled},
    the Casimir preserves its functional form relative to the undeformed case:
    \begin{equation}
        C_{h}= Z_{+}Z_{-}+ \frac{H}{2}\left(\frac{H}{2}- \mathbf{I}\right). \label{hcas}
    \end{equation}
    According to \cite{Aizawa1997} for finite-dimensional
    irreducible representations, the generators act on the standard basis
    $|j, m\rangle$ as follows:
    \begin{align}
        H |j, m\rangle    & = 2m |j, m\rangle, \nonumber   \\
        Z_{+}|j, m\rangle & = |j, m+1\rangle, \nonumber    \\
        Z_{-}|j, m\rangle & = (j+m)(j-m+1) |j, m-1\rangle.
    \end{align}
 Although this action seems to mimic the classical one, the structure of the
    coproduct $\Delta_{h}$ is what distinguishes the system. The Jordanian deformation breaks hermiticity of the Hamiltonian because it introduces an inherently asymmetric coproduct for $Z_+, Z_-$ (Eq.\eqref{eq:coproduct_j}) that is fundamentally incompatible with the standard involution ($*$-structure) of the $\mathfrak{sl}(2, \mathbb{R})$ algebra (which defines the adjoint relation $Z_\pm^\dagger=Z_\mp$ ). To build a multi-site or many-body Hamiltonian, one must iteratively apply this asymmetric coproduct (Eq.\eqref{fl}). Because the $h$-deformed Hamiltonian is non-Hermitian, one loses the unique orthonormal basis guaranteed by the spectral theorem, introducing an inherent arbitrariness into the eigenvectors. To resolve this geometric ambiguity, one can leverage the underlying algebraic symmetry: constructing a basis using h-Clebsch--Gordan coefficients uniquely privileges a specific set of states letting the quantum group's structure naturally dictate the physical basis of the system. 
    Indeed, as can be seen in Appendix \ref{app1:sec2}, the eigenvectors of the $h$-KS model are smooth $h$-
deformations of the usual undeformed Dicke states that include qubit states with $k$ extra excitations
endowed with weights of the type $h^k$. Moreover, the spectrum remains invariant because the total Hamiltonian retains a triangular block structure in the uncoupled basis given by the $h$-Clebsch--Gordan coefficients, meaning the deformation alters only the off-diagonal coupling without shifting the bare energy levels.  Consequently, the energy levels of the $h$-KS model are identical to those
    of the undeformed Kittel--Shore model for any value of $h$. This
    distinguishes the Jordanian case from the $q$-KS model, where the energy levels
    are explicitly shifted by the $q$-deformation.

    Spectral invariance has physical implications. From a thermodynamic
    perspective, the partition function $Z = \text{Tr}(e^{-\beta H})$ and all derived
    observables—such as specific heat, internal energy, and magnetic susceptibility—remain
    unaffected by the $h$-deformation. Thus, the $h$-KS model is thermodynamically
    indistinguishable from the classical Kittel--Shore system.

    However, the $h$-deformation is not physically trivial. While the
    eigenvalues are preserved, the eigenvectors are deformed, as seen in Eq. (\ref{program1})
    and Appendix \ref{app1:sec2}. These ``Jordanian eigenstates'' are no longer
    the standard states but rather complex superpositions that break the
    exchange symmetry of the computational basis. This makes the $h$-KS model a highly
    interesting framework for quantum information theory. The deformation
    parameter $h$ modulates the quantum correlations without changing the energy
    of the states, providing a unique mechanism to control the correlation structure
    of a qubit network while keeping its thermal properties constant.

    Identifying the specific physical system is vital for a meaningful interpretation
    of the results, as it allows the deformation parameters $q$ and $h$ to be viewed not merely as abstract mathematical parameters, but as physical coupling constants that induce anisotropy and modulate the correlation structure within a superintegrable qubit network. This physical modulation manifests as a redistribution of the state weights in both deformed cases, with the h-deformation uniquely introducing novel quantum states that are fundamentally absent in the undeformed model. From a physical point of view, these deformations allow us to parameterize experimental non-idealities like anisotropic noise or environmental coupling, giving the fidelities in Sec.~\ref{sec:entropy} a distinct phenomenological relevance. By revealing how breaking or deforming the underlying symmetry reshapes the distribution and stability of quantum information across the eigenstates, the deformed Kittel--Shore model provides a concrete framework for entanglement engineering in next-generation quantum devices.

    \section{Comparison with experimental results}
    \label{sec:comparison}  Before proceeding with a detailed characterization of
the fidelities, it is interesting to establish a connection between the
deformed eigenstates and the states typically targeted in quantum
information experiments. This section presents a comparison between the experimental
results and the theoretical predictions obtained using the $q$- and $h$-deformation
models. To demonstrate the suitability of both deformations with respect to their
potential to describe experimental output, we focus on the results of two
papers, one for each deformation.

The deformed states considered in our approach are constructed algebraically by
coupling spin-$1/2$ representations through the corresponding deformed
Clebsch--Gordan coefficients. On the experimental side, the target states are prepared by means of quantum
circuits specifically designed for Dicke-type states; in
particular, when mitigated data are provided, readout errors are reduced by
calibrating the measurement response of the device and correcting the observed
probability distributions before the comparison with theory.

As previously stated in the above Section, the $q$-deformation weights the states in a $q$-dependent
manner without introducing any new state. However, $h$-deformation
introduces new states proportional to $h$ to the original states. Therefore,
these deformations can be applied to different experimental situations, depending on whether
new states appear.

We use $q$-deformation to describe the states of \cite{aktar2022divide} and $h$-deformation
for those of \cite{cruz2019efficient}. This, as we will see, is motivated by
the fact that the former states present a negligible contribution to the new
states, whereas in the latter, a significant contribution appears.

The choice of these two experimental works is guided by the availability and
compatibility of the reported data. For the present comparison, it is not
sufficient that a given work prepares a multipartite entangled state; the
experimental probability distribution, or enough information to reconstruct it,
must be explicitly provided, and the target states must be compatible with the
families of states generated by the deformed algebraic construction. To the best of our knowledge, among the available experimental implementations
with sufficiently detailed reported data, the results reported in
\cite{aktar2022divide} and \cite{cruz2019efficient} provide a particularly
suitable benchmark, since they include experimentally accessible data for
Dicke-type states and allow for a direct comparison with
the theoretical probabilities predicted by the \(q\)- and \(h\)-deformed
models. 

    \subsection{Comparison of experimental results with \texorpdfstring{$q$}{q}-deformation}

\begin{figure}[H]
        \centering
        \begin{minipage}[t]{0.48\linewidth}
            \centering
            \includegraphics[width=\linewidth]{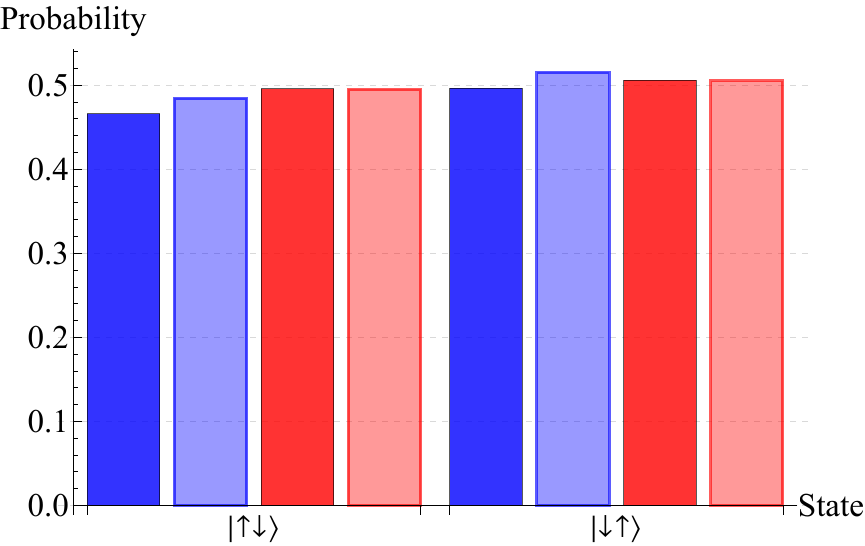}
        \end{minipage}
        \hfill
        \begin{minipage}[t]{0.48\linewidth}
            \centering
            \includegraphics[width=\linewidth]{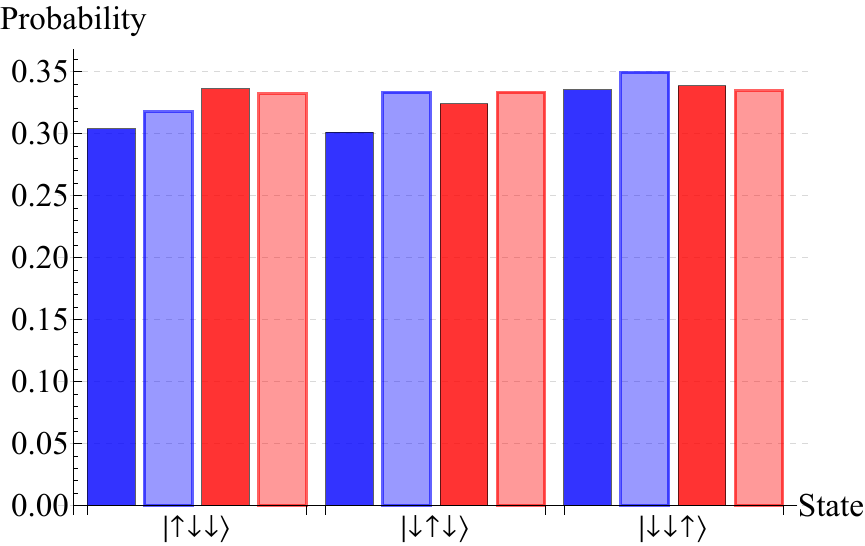}
        \end{minipage}
        \caption{Comparison of experimental data\cite{aktar2022divide} and theoretical predictions for $\ket{D_2^0}_{q}$ (left) and $\ket{D_3^{-1/2}}_{q}$ (right). Dark blue (dark red) indicates the raw device (error-mitigated) experimental probabilities, while light blue (light red) shows the corresponding theoretical values from the $q$-deformation.}
        \label{fig:wq23}
    \end{figure}

\begin{figure}[H]
        \centering
        \begin{minipage}[t]{0.48\linewidth}
            \centering
            \includegraphics[width=\linewidth]{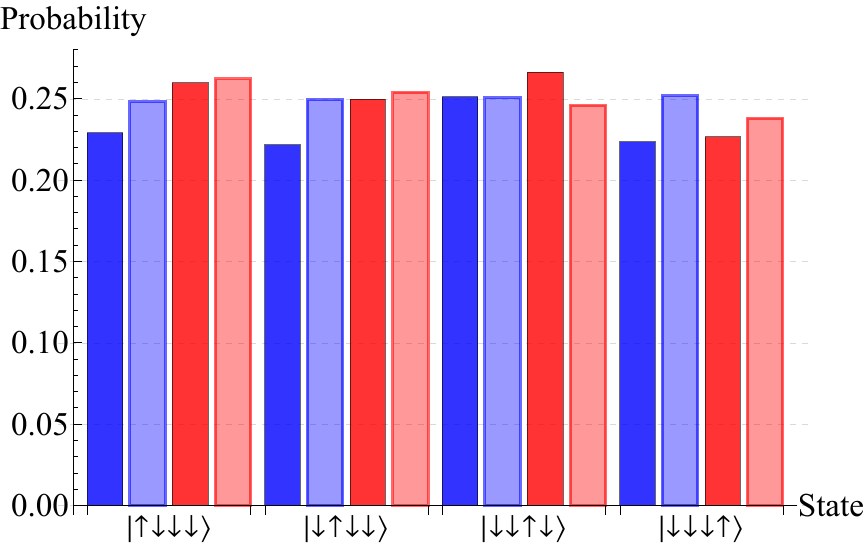}
        \end{minipage}
        \hfill
        \begin{minipage}[t]{0.48\linewidth}
            \centering
            \includegraphics[width=\linewidth]{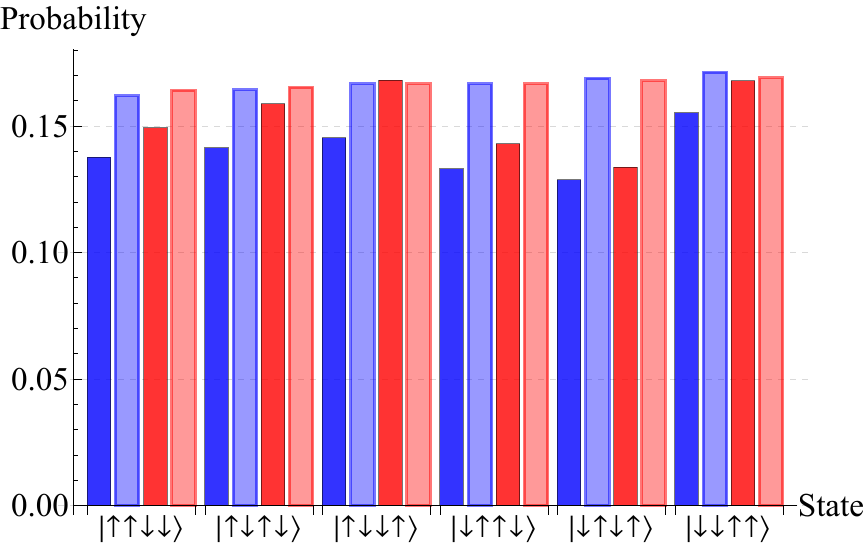}
        \end{minipage}
        \caption{Comparison of experimental data\cite{aktar2022divide} and theoretical predictions for $\ket{D_4^{-1}}_{q}$ (left) and $\ket{D_4^{0}}_{q}$ (right). Dark blue (dark red) indicates the raw device (error-mitigated) experimental probabilities, while light blue (light red) shows the corresponding theoretical values from the $q$-deformation. }
        \label{fig:wq4}
    \end{figure}

    This subsection presents a detailed comparison between the experimental results
    of a \textit{IBM Q Montreal device} \cite{aktar2022divide} (both for raw states directly obtained from the device and for error-mitigated states) and our theoretical predictions using $q$-deformation
    models.  Specifically, our analysis focuses on a direct comparison between the states $\ket{D_2^0}$,
    $\ket{D_3^{-1/2}}$, $\ket{D_4^{-1}}$, and $\ket{D_4^{0}}$ studied in their paper and the corresponding deformed states obtained from our theoretical framework $\ket{D_2^0}_q$,
    $\ket{D_3^{-1/2}}_q$, $\ket{D_4^{-1}}_q$, and $\ket{D_4^{0}}_q$.   Using the data available in \cite{aktar2022divide}, we obtain the optimal value
of \(q\) by a least-squares fit of the probability of each component of the target states in the computational basis
\begin{equation}
\mathcal{B}_N=
\left\{
\ket{s_1s_2\cdots s_N}\;:\;
s_\ell\in\{\uparrow,\downarrow\},\ \ell=1,\ldots,N
\right\},
\label{eq22}
\end{equation}
where
\begin{equation}
\ket{s_1s_2\cdots s_N}
=
\ket{s_1}\otimes\ket{s_2}\otimes\cdots\otimes\ket{s_N}.
\end{equation}
Results are shown in Figs. \ref{fig:wq23}, \ref{fig:wq4} as well as in
    Tables \ref{table:w2}, \ref{table:w3}, \ref{table:w4}, and \ref{table:w42}.  In all the tables, the columns specify the considered quantum state, the experimental probability directly measured from the device (Device), the theoretical probability from the $q$-deformation fitted to the device data ($q$-Device), the probability after applying error mitigation (Mitigated), and the theoretical probability from the $q$-deformation fitted to the mitigated data ($q$-Mitigated). 

    \begin{table}[h!]
        \centering
        \begin{tabular}{lcccc}
            \hline
            State                       & Device & $q$-Device & Mitigated & $q$-Mitigated \\
            \hline
            $\ket{\uparrow \downarrow}$ & 0.4658 & 0.4846     & 0.4956    & 0.4951        \\
            $\ket{\downarrow \uparrow}$ & 0.4967 & 0.5155     & 0.5055    & 0.5050        \\
            \hline
            Value of $q$                & ---    & 1.0638     & ---       & 1.0200        \\
            \hline
        \end{tabular}
        \caption{Experimental data of the $\ket{D_2^0}$ state versus the theoretical $\ket{D_2^0}_q$ in Fig.~\ref{fig:wq23} (left).}
        \label{table:w2}
    \end{table}

    \begin{table}[h!]
        \centering
        \begin{tabular}{lcccc}
            \hline
            State                                  & Device & $q$-Device & Mitigated & $q$-Mitigated \\
            \hline
            $\ket{\uparrow \downarrow \downarrow}$ & 0.3039 & 0.3174     & 0.3362    & 0.3322        \\
            $\ket{\downarrow \uparrow \downarrow}$ & 0.3009 & 0.3331     & 0.3239    & 0.3333        \\
            $\ket{\downarrow \downarrow \uparrow}$ & 0.3354 & 0.3495     & 0.3385    & 0.3345        \\
            \hline
            Value of $q$                           & ---    & 1.0493     & ---       & 1.0035        \\
            \hline
        \end{tabular}
        \caption{Experimental data of the $\ket{D_3^{-1/2}}$ state versus the theoretical $\ket{D_3^{-1/2}}_q$ in Fig.~\ref{fig:wq23} (right).}
        \label{table:w3}
    \end{table}

    \begin{table}[h!]
        \centering
        \begin{tabular}{lcccc}
            \hline
            State                                             & Device & $q$-Device & Mitigated & $q$-Mitigated \\
            \hline
            $\ket{\uparrow \downarrow \downarrow \downarrow}$ & 0.2291 & 0.2481     & 0.2597    & 0.2622        \\
            $\ket{\downarrow \uparrow \downarrow \downarrow}$ & 0.2218 & 0.2494     & 0.2495    & 0.2539        \\
            $\ket{\downarrow \downarrow \uparrow \downarrow}$ & 0.2513 & 0.2506     & 0.2663    & 0.2459        \\
            $\ket{\downarrow \downarrow \downarrow \uparrow}$ & 0.2236 & 0.2519     & 0.2267    & 0.2381        \\
            \hline
            Value of $q$                                      & ---    & 1.0051     & ---       & 0.9684        \\
            \hline
        \end{tabular}
        \caption{Experimental data of the $\ket{D_4^{-1}}$ state versus the theoretical $\ket{D_4^{-1}}_q$ in Fig.~\ref{fig:wq4} (left).}
        \label{table:w4}
    \end{table}

    \begin{table}[h!]
        \centering
        \begin{tabular}{lcccc}
            \hline
            State                                           & Device & $q$-Device & Mitigated & $q$-Mitigated \\
            \hline
            $\ket{\uparrow \uparrow \downarrow \downarrow}$ & 0.1377 & 0.1621     & 0.1493    & 0.1642        \\
            $\ket{\uparrow \downarrow \uparrow \downarrow}$ & 0.1414 & 0.1643     & 0.1589    & 0.1654        \\
            $\ket{\uparrow \downarrow \downarrow \uparrow}$ & 0.1454 & 0.1666     & 0.1682    & 0.1667        \\
            $\ket{\downarrow \uparrow \uparrow \downarrow}$ & 0.1331 & 0.1666     & 0.1430    & 0.1667        \\
            $\ket{\downarrow \uparrow \downarrow \uparrow}$ & 0.1288 & 0.1690     & 0.1338    & 0.1679        \\
            $\ket{\downarrow \downarrow \uparrow \uparrow}$ & 0.1553 & 0.1713     & 0.1679    & 0.1691        \\
            \hline
            Value of $q$                                    & ---    & 1.0140     & ---       & 1.0074        \\
            \hline
        \end{tabular}
        \caption{Experimental data of the $\ket{D_4^{0}}$ state versus the theoretical $\ket{D_4^{0}}_q$ in Fig.~\ref{fig:wq4} (right).}
        \label{table:w42}
    \end{table}

    It can be seen that among all possible values, the best fit to the experimental data is consistently achieved for $q$ close to 1 (with the maximum and minimum fitted values being 1.0638 and 0.9684, respectively). Moreover, our $q$-deformed model exhibits a significantly better agreement with the error-mitigated states than with the raw states directly obtained from the device because the latter are not properly normalized within the
computational subspace considered in the fit. Since the theoretical
\(q\)-deformed probabilities form a normalized distribution, measurement-error
mitigation restores the total probability closer to one. This indicates that the deformation successfully captures the idealized target state once experimental noise is filtered out.

    \subsection{Comparison of experimental results with \texorpdfstring{$h$}{h}-deformation}
\begin{figure}[H]
        \centering
        \includegraphics[scale=0.8]{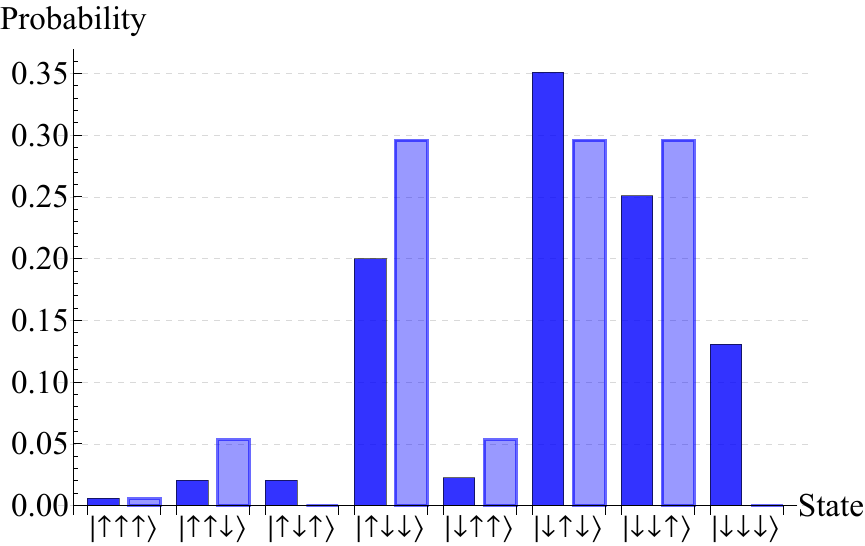}
        \caption{Comparison of experimental data\cite{cruz2019efficient} and theoretical predictions for $\ket{D_3^{-1}}_{h}$. Dark blue indicates the raw device experimental probabilities while light blue shows the corresponding theoretical values from the $h$-deformation.}
        \label{fig:w32}
    \end{figure}

\begin{figure}[H]
        \centering
        \includegraphics[scale=0.7, trim=90 1 1 1, clip]{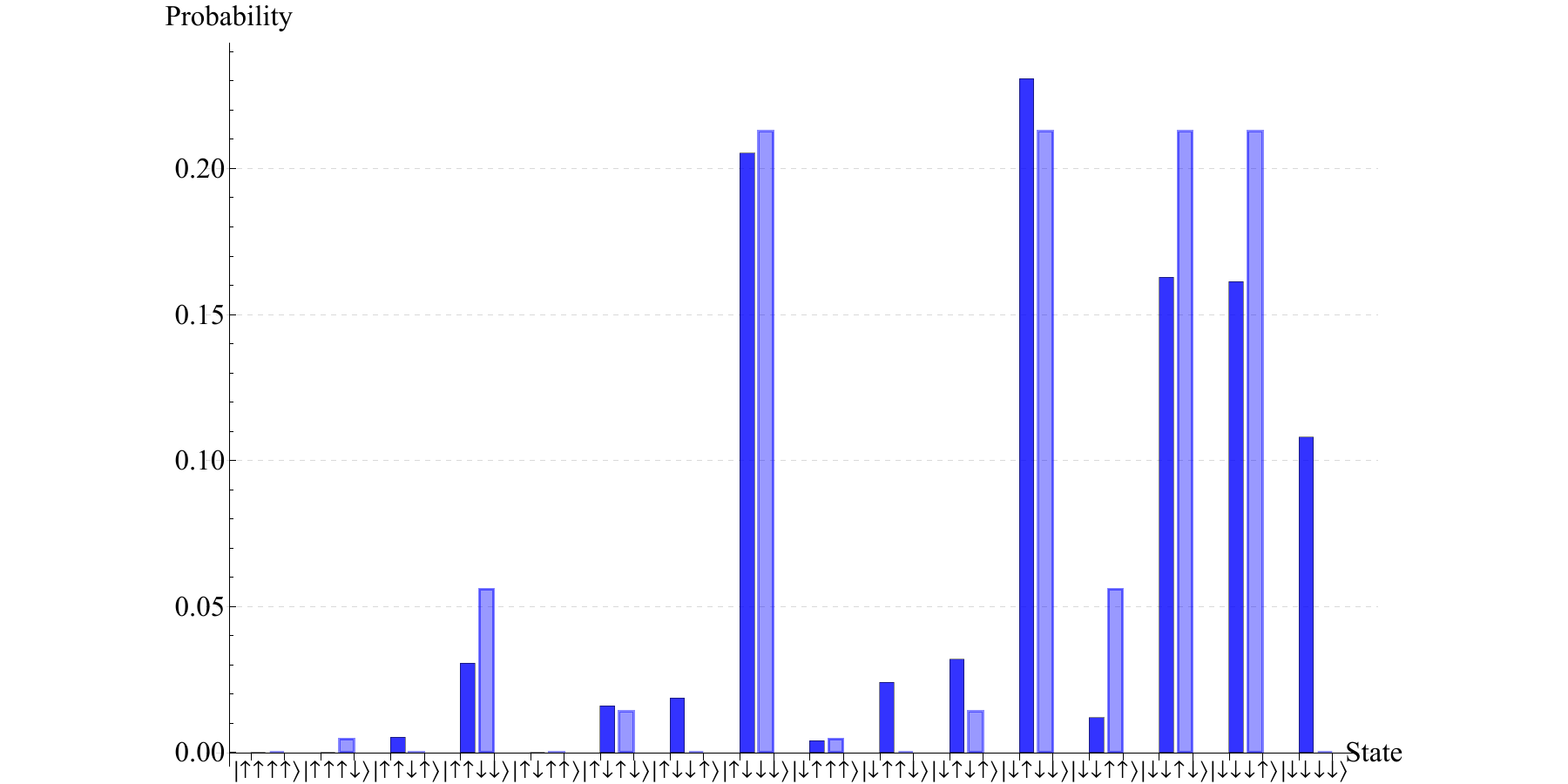}
        \caption{Comparison of experimental data\cite{cruz2019efficient} and theoretical predictions for $\ket{D_4^{-2}}_{h}$. Dark blue indicates the raw device experimental probabilities while light blue shows the corresponding theoretical values from the $h$-deformation.}
        \label{fig:w43}
    \end{figure}

    We will discuss the experimental results for the states $\ket{D_3^{-1}}$ and
    $\ket{D_4^{-2}}$ obtained in \cite{cruz2019efficient} through the
    \textit{IBM Q platform} on the Internet. In this case, because new states emerge that differ from those of the undeformed model, the use of the $h$-deformation allows for a
    better description.  Again, the optimal value of $h$ is obtained by comparing the experimental data with our theoretical predictions and performing a least-squares adjustment of the probability of each component of the target states in the computational basis. The results are shown in Figs. \ref{fig:w32} and \ref{fig:w43}, as well as in Tables \ref{table:w32} and \ref{table:w43}. In all tables, the columns detail the considered quantum state (State), the experimentally obtained probability (Exp.), and the corresponding
        theoretical probability derived from the $h$-deformation model ($h$-def.). Note that in \cite{cruz2019efficient} only raw, unmitigated device data are reported.

    \begin{table}[h!]
        \centering
        \begin{tabular}{lcc}
            \hline
            State                                  & Exp. & $h$-def \\
            \hline
            $\ket{\uparrow\uparrow\uparrow}$       & 0.0061       & 0.0054          \\
            $\ket{\uparrow\uparrow\downarrow}$     & 0.0204       & 0.0533          \\
            $\ket{\uparrow\downarrow\uparrow}$     & 0.0204       & 0.0000          \\
            $\ket{\uparrow\downarrow\downarrow}$   & 0.2000       & 0.2960          \\
            $\ket{\downarrow\uparrow\uparrow}$     & 0.0225       & 0.0533          \\
            $\ket{\downarrow\uparrow\downarrow}$   & 0.3510       & 0.2960          \\
            $\ket{\downarrow\downarrow\uparrow}$   & 0.2510       & 0.2960          \\
            $\ket{\downarrow\downarrow\downarrow}$ & 0.1306       & 0.0000          \\
            \hline
        \end{tabular}
        \caption{Experimental data of the $\ket{D_3^{-1}}$ state versus the theoretical $\ket{D_3^{-1}}_h$ in Fig.~\ref{fig:w32} for $h=0.4243$.}
        \label{table:w32}
    \end{table}

    \begin{table}[h!]
        \centering
        \begin{minipage}[t]{0.48\linewidth}
            \centering
            \hspace{2cm}
            \begin{tabular}{lcc}
                \hline
                State                                          & Exp.   & $h$-def. \\
                \hline
                $\ket{\uparrow\uparrow\uparrow\uparrow}$       & 0.0000 & 0.0000   \\
                $\ket{\uparrow\uparrow\uparrow\downarrow}$     & 0.0000 & 0.0047   \\
                $\ket{\uparrow\uparrow\downarrow\uparrow}$     & 0.0053 & 0.0001   \\
                $\ket{\uparrow\uparrow\downarrow\downarrow}$   & 0.0307 & 0.0559   \\
                $\ket{\uparrow\downarrow\uparrow\uparrow}$     & 0.0000 & 0.0001   \\
                $\ket{\uparrow\downarrow\uparrow\downarrow}$   & 0.0160 & 0.0140   \\
                $\ket{\uparrow\downarrow\downarrow\uparrow}$   & 0.0187 & 0.0000   \\
                $\ket{\uparrow\downarrow\downarrow\downarrow}$ & 0.2053 & 0.2127   \\
                \hline
            \end{tabular}
        \end{minipage}
        \hfill
        \begin{minipage}[t]{0.48\linewidth}
            \hspace{-2cm}
            \centering
            \begin{tabular}{lcc}
                \hline
                State                                            & Exp.   & $h$-def. \\
                \hline
                $\ket{\downarrow\uparrow\uparrow\uparrow}$       & 0.0040 & 0.0047   \\
                $\ket{\downarrow\uparrow\uparrow\downarrow}$     & 0.0240 & 0.0000   \\
                $\ket{\downarrow\uparrow\downarrow\uparrow}$     & 0.0320 & 0.0140   \\
                $\ket{\downarrow\uparrow\downarrow\downarrow}$   & 0.2307 & 0.2127   \\
                $\ket{\downarrow\downarrow\uparrow\uparrow}$     & 0.0120 & 0.0559   \\
                $\ket{\downarrow\downarrow\uparrow\downarrow}$   & 0.1627 & 0.2127   \\
                $\ket{\downarrow\downarrow\downarrow\uparrow}$   & 0.1613 & 0.2127   \\
                $\ket{\downarrow\downarrow\downarrow\downarrow}$ & 0.1080 & 0.0000   \\
                \hline
            \end{tabular}
        \end{minipage}
        \vspace{0.3em}

        \caption{Experimental data of the $\ket{D_4^{-2}}$ state versus the theoretical $\ket{D_4^{-2}}_h$ in Fig.~\ref{fig:w43} for $h=0.2564$. }
        \label{table:w43}
    \end{table} 

Notice that in the theoretical model the \(h\)-deformed states deviate more significantly
from the undeformed counterparts than in the \(q\)-deformation. This wider
discrepancy arises because the \(h\)-deformed states include specific
computational-basis components with non-vanishing probability, despite being
completely absent in the undeformed case. Figs.~\ref{fig:w32} and
\ref{fig:w43} show this fact. However, one can also see that some of the computational
basis components observed experimentally are not present in the theoretical
\(h\)-deformed states although the overall structure of the theoretical distributions still
resembles the experimental ones. These differences are mainly related to the relative weights of the
allowed components. In particular, the \(h\)-deformation assigns the same
probability to one-excitation terms, whereas the experimental data
show an asymmetric distribution among them. This indicates that the
\(h\)-deformation captures the appearance of additional components, but not the
full redistribution of weights observed experimentally. Therefore, a more
accurate description of these data may require a hybrid \((q,h)\)-deformation,
since the observed deviations combine two effects: the appearance of additional
computational-basis components, naturally captured by the \(h\)-deformation,
and a redistribution of the weights of the expected components, which is
characteristic of the \(q\)-deformation.

    \section{Fidelities of deformed states}
    \label{sec:entropy} The aim of this section is to study the fidelities of
    the $q$- and $h$-deformed states presented in \cite{Ballesteros:2025cia,Ballesteros:2025dbv}
    and \cite{ballesteros2025entangled}, respectively, with their corresponding undeformed
    counterparts.

    \subsection{Deformed states with two qubits}
    \label{sec:2q}

\begin{figure}[H]
        \centering
        \begin{minipage}[t]{0.48\linewidth}
            \centering
            \includegraphics[width=\linewidth]{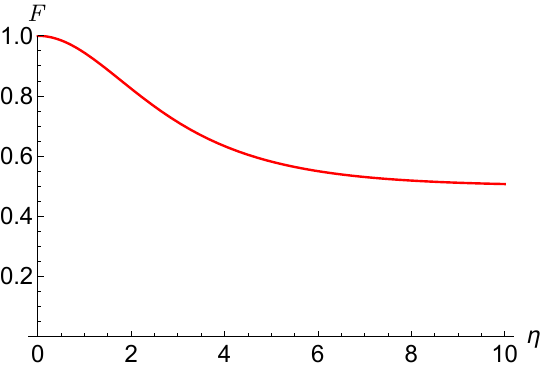}
        \end{minipage}
        \hfill
        \begin{minipage}[t]{0.48\linewidth}
            \centering
            \includegraphics[width=\linewidth]{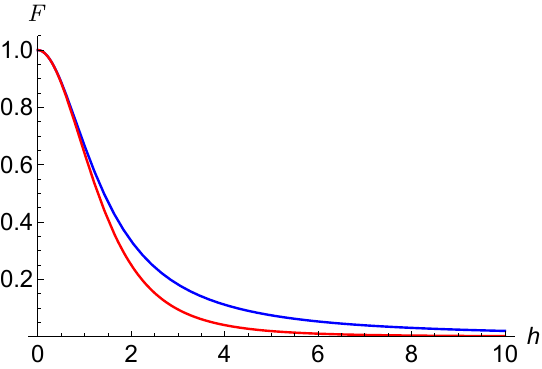}
        \end{minipage}
        \caption{\small Left: Fidelity between the $q$-deformed state $\ket{D_2^{0}}
        _{q}$ (red) (the same for $\ket{M_2^{0}}_{q}$) and its corresponding undeformed
        state as a function of the deformation parameter. Right: Fidelity between
        the $h$-deformed states $\ket{M_2^0}_{h}$ (blue) and $\ket{D_2^{-2}}_{h}$(red)
        and their corresponding undeformed states as a function of the
        deformation parameter.}
        \label{fig:fid2}
    \end{figure}

    The fidelity of the $q$-deformed states $\lvert M_{2}^{0}\rangle_{q}$ and
    $\lvert D_{2}^{0}\rangle_{q}$ is given by
    \begin{equation}
        F(\lvert M_{2}^{0}\rangle_{q=1},\lvert M_{2}^{0}\rangle_{q})=F(\lvert D_{2}
        ^{0}\rangle_{q=1},\lvert D_{2}^{0}\rangle_{q})=\frac{1}{2}+\frac{\sqrt{q}}{1+q}
        , \label{eq:fid}
    \end{equation}
    and is shown  in Fig.~\ref{fig:fid2} (left). For
    representation convenience, we consider $q=e^{\eta}$. As expected, the fidelity
    equals one at $q=1$ ($\eta=0$), where both states coincide with their
    undeformed counterparts. As $\eta$ increases, the fidelity decreases
    monotonically, reflecting the gradual deviation from the original state caused by the standard deformation.
Notice that the state fidelity between the deformed and undeformed configurations exhibits an asymptotic saturation rather than decaying to zero in the limit $q\rightarrow \infty$. This behavior is deeply rooted in the Kashiwara crystal limit\footnote{Although Kashiwara originally studied the limit $q \to 0$, evaluating the asymptotic fidelity in the limit $q \to \infty$ is mathematically equivalent to the limit $q \to 0$. This equivalence arises from the strict invariance of the $q$-numbers under the involution $q \leftrightarrow q^{-1}$. Consequently, the fidelity yields identical values in both asymptotic regimes, as the system effectively ``freezes'' into a singular, highly localized state.} \cite{kashiwara1991crystal}. As $q$ approaches this critical limit, the continuous algebraic structure of the representation transitions into a rigid, combinatorially well-defined framework known as a crystal graph. Within a finite-dimensional Hilbert space—such as the spin-1/2 representation—the deformed states do not escape into orthogonal regions of the state space; instead, they smoothly converge toward the rigid basis vectors of the crystal lattice. This topological preservation bounds the geometric distance between the states, preventing total orthogonality and explaining the non-vanishing saturation of the quantum fidelity.

    The fidelities between the $h$-deformed states and their corresponding
    undeformed states for two qubits are plotted in Fig.~\ref{fig:fid2} (right) and
    can be written as:
    \begin{eqnarray}
        F( \ket{D_2^{-2}}_{h=0}, \ket{D_2^{-2}}_h)=
        \frac{16}{\left(h^{2}+4\right)^{2}}, \qquad F(\ket{M_2^{0}}_{h=0},\ket{M_2^{0}}_h)=
        \frac{2}{2+h^{2}}.
    \end{eqnarray}
    
    As also observed in the $q$-deformation case, the fidelity $F$ for both $\lvert M_{2}^{0}\rangle_{h}$ and $\lvert D_{2}^{-2}\rangle_{h}$ starts at $1$ when $h = 0$, where the states coincide exactly in the absence of deformation. As $h$ increases, the fidelity decreases steadily toward zero, indicating that larger deformations push each deformed state farther away from its corresponding undeformed configuration. Notably, $\lvert D_{2}^{-2}\rangle_{h}$ exhibits a higher sensitivity to the deformation than $\lvert M_{2}^{0}\rangle_{h}$. By the time $h \approx 6$, both fidelities drop below $0.1$, rendering the deformed states nearly orthogonal to their original versions. 
    The fact that $\lvert D_{2}^{-2}\rangle_{h}$ loses fidelity faster suggests that its underlying algebraic or tensor product structure is more drastically reshaped by the $h$-deformation.

The rapid collapse stands in sharp contrast to the standard $q$-deformed case (Fig.~\ref{fig:fid2}, left), where the fidelity preserves a residual fraction and saturates at a non-zero value. This fundamental discrepancy is deeply rooted in the algebraic nature of each mechanism. The standard $q$-deformation smoothly guides the states toward the rigid, combinatorially well-defined framework of the Kashiwara crystal limit as the parameter approaches infinity. Within a finite-dimensional Hilbert space, this topological preservation prevents total orthogonality. Conversely, this saturation effect is entirely absent in the case of the Jordanian deformation. While the standard $q$-deformation scales the weights of the representation systematically, the $h$-deformation introduces non-symmetric terms involving nilpotent generators. As the deformation parameter increases, these nilpotent contributions dominate the transformation of the states, rapidly suppressing their overlap with the original configuration until complete orthogonality is reached. Consequently, while the $q$-deformation acts as a smoother mechanism that maintains a stable fidelity fraction, the Jordanian deformation produces a highly disruptive modification, allowing one to effectively distinguish the deformed state from its original ancestor with near certainty.

    \subsection{Deformed states with three qubits}
    \label{sec:3q}

\begin{figure}[H]
        \centering
        \begin{minipage}[t]{0.48\linewidth}
            \centering
            \includegraphics[width=\linewidth]{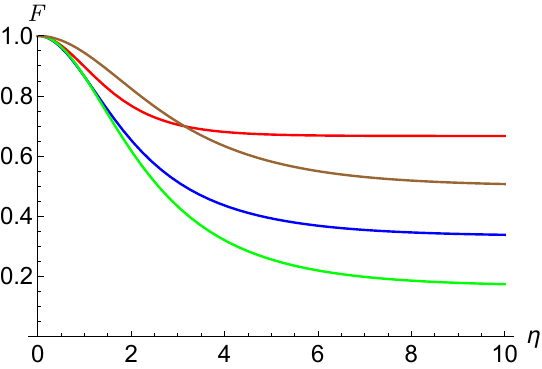}
        \end{minipage}
        \hfill
        \begin{minipage}[t]{0.48\linewidth}
            \centering
            \includegraphics[width=\linewidth]{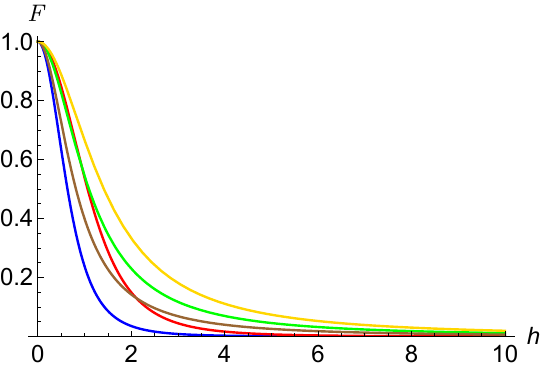}
        \end{minipage}
        \caption{\small Fidelity between the deformed states and their corresponding undeformed states
        as a function of the deformation parameter. Left: Cases involving $\ket{D_3^{\pm1/2}}
        _{q}$ (blue), $\ket{M_3^{-1/2}}_{q}$ (green), $\ket{M_3^{1/2}}_{q}$ (red)
        and $\ket{V_3^{\pm1/2}}_{q}$ (brown). Right: Cases involving $\ket{D_3^{-3}}_{h}$ (blue), $\ket{D_3^{-1}}_{h}$(red),
        $\ket{M_3^{-1}}_{h}$ (green), $\ket{M_3^{1}}_{h}$(brown) and
        $\ket{V_3^{\pm1}}_{h}$(yellow).}
        \label{fig:fid3}
    \end{figure}

    When comparing the $q$-deformed states with $3$ qubits with their
    corresponding undeformed counterparts, we find the following fidelities, shown in Fig. \ref{fig:fid3} (left) as a function of $\eta$:
    \begin{align}
         & F(|D^{-1/2}_{3}\rangle_{q=1},|D^{-1/2}_{3}\rangle_{q})=F(|D^{1/2}_{3}\rangle_{q=1},|D^{1/2}_{3}\rangle_{q})=\frac{1+\sqrt{q}+q}{3(1-\sqrt{q}+q)}, \notag \\
         & F(|M^{-1/2}_{3}\rangle_{q=1},|M^{-1/2}_{3}\rangle_{q})=\frac{(q^{3/2}+ 3q + 2)^{2}}{6(q+1)(q^{2}+q+1)}, \notag                                           \\
         & F(|M^{1/2}_{3}\rangle_{q=1},|M^{1/2}_{3}\rangle_{q})=\frac{\left((2 q+3) \sqrt{q}+1\right)^{2}}{6 (q+1) \left(q^{2}+q+1\right)}, \notag                  \\
         & F(|V^{-1/2}_{3}\rangle_{q=1},|V^{-1/2}_{3}\rangle_{q})=F(|V^{1/2}_{3}\rangle_{q=1},|V^{1/2}_{3}\rangle_{q})=\frac{q+2 \sqrt{q}+1}{2 q+2}.
    \end{align}
    Once again, the fidelity starts at $1$ in the undeformed limit $\eta=0$ and subsequently decreases as the deformation parameter increases. The impact of
    deformation varies slightly among different states. Regarding the specific configurations, the fidelities of $\ket{D_3^{\pm1/2}}_{q}$ experience a steady decline before leveling off near $0.35$ at large $\eta$, indicating that a significant residual overlap survives the deformation. A comparable trend is found for $\ket{V_3^{\pm1/2}}_{q}$, which stabilize around $0.50$ and confirm a similar vulnerability to the $q$-mechanism. The remaining cases, however, deviate noticeably from this average behavior: while $\ket{M_3^{-1/2}}_{q}$ proves highly sensitive—dropping rapidly toward an asymptote of approximately $0.20$—the $\ket{M_3^{1/2}}_{q}$ state emerges as the most resilient configuration, decaying slowly to maintain a strong structural similarity to its original form at $0.67$. As synthesized in Fig.~\ref{fig:fid3} (left), the standard $q$-deformation induces distinct, parameter-dependent footprints based on the algebraic structure of each state, yet all share a common saturation profile. As discussed previously, this asymptotic freezing as $\eta \to \infty$ is directly linked to the Kashiwara crystal limit, where the quantum state simplifies into a single dominant basis vector, forcing the fidelity to lock into a constant, non-vanishing value.

    For the $3$-qubit states in the $h$-deformation case with $3$ qubits, the following fidelities are found
    \begin{eqnarray}
        &&F( \ket{D_3^{-3}}_{h=0}, \ket{D_3^{-3}}_h)=\frac{16}{19 h^{4}+32 h^{2}+16},
        \qquad F( \ket{D_3^{-1}}_{h=0}, \ket{D_3^{-1}}_h)=\frac{48}{9 h^{4}+32 h^{2}+48},\nonumber\\
        &&F( \ket{M_3^{-1}}_{h=0}, \ket{M_3^{-1}}_h)=\frac{6}{5 h^{2}+6}, \qquad
        \qquad \qquad \, F( \ket{M_3^{1}}_{h=0}, \ket{M_3^{1}}_h)=\frac{2}{3 h^{2}+2}
        , \nonumber\\
        &&F( \ket{V_3^{-1}}_{h=0}, \ket{V_3^{-1}}_h)= F( \ket{V_3^{1}}_{h=0}, \ket{V_3^{1}}_h)=\frac{2}{h^{2}+2}.
    \end{eqnarray}
    The results are shown in Fig.~\ref{fig:fid3} (right). The behavior is totally analogous to that of two-qubit states: the fidelity decreases 
    with $h$  until it reaches zero for all states.  The rate at which fidelity decreases depends on the state: $|D_{3}
    ^{-3}\rangle_{h}$ and $|D_{3}^{-1}\rangle_{h}$ lose fidelity more rapidly for
    increasing $h$ compared to $|V_{3}^{-1}\rangle_{h}$ and $|M_{3}^{-1}\rangle_{h}$,
    while $|M_{3}^{1}\rangle_{h}$ shows an intermediate behavior.

        \subsection{Deformed states with four qubits}
    \label{sec:4q}

\begin{figure}[H]
        \centering
        \begin{minipage}[t]{0.48\linewidth}
            \centering
            \includegraphics[width=\linewidth]{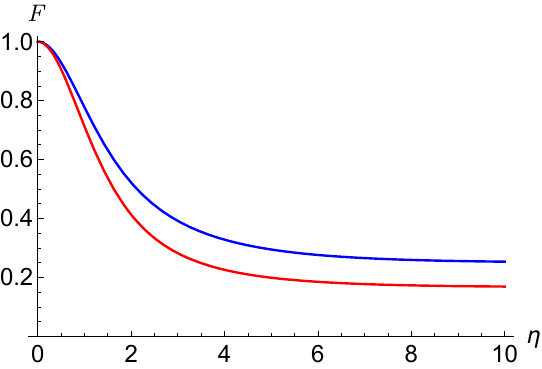}
        \end{minipage}
        \hfill
        \begin{minipage}[t]{0.48\linewidth}
            \centering
            \includegraphics[width=\linewidth]{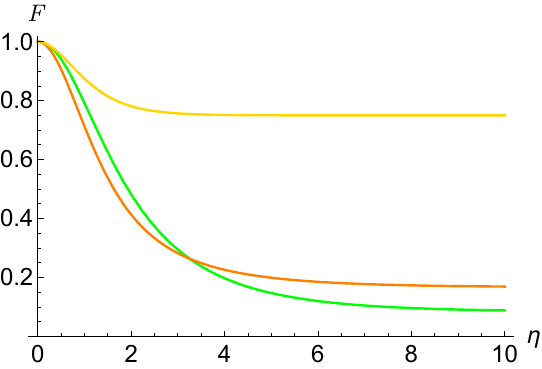}
        \end{minipage}
        \caption{\small Fidelity between the $q$-deformed states and their corresponding undeformed states as a function of the deformation parameter. Left: Cases involving $\ket{D_4^{\pm 1}}_{q}$ (blue) and $\ket{D_4^{0}}_{q}$ (red). Right: Cases involving $\ket{M_4^{-1}}_{q}$ (green), $\ket{M_4^{0}}_{q}$ (orange) and $\ket{M_4^{1}}_{q}$ (yellow). Notice that $F(|D^{0}_{4}\rangle_{q=1},|D^{0}_{4}\rangle_{q}) =F(|M^{0}_{4}\rangle_{q=1},|M^{0}_{4}\rangle_{q})$.}
        \label{fig:fiq41}
    \end{figure}

\begin{figure}[H]
        \centering
        \begin{minipage}[t]{0.48\linewidth}
            \centering
            \includegraphics[width=\linewidth]{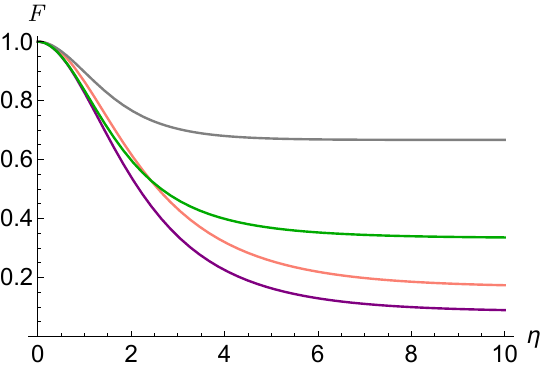}
        \end{minipage}
        \hfill
        \begin{minipage}[t]{0.48\linewidth}
            \centering
            \includegraphics[width=\linewidth]{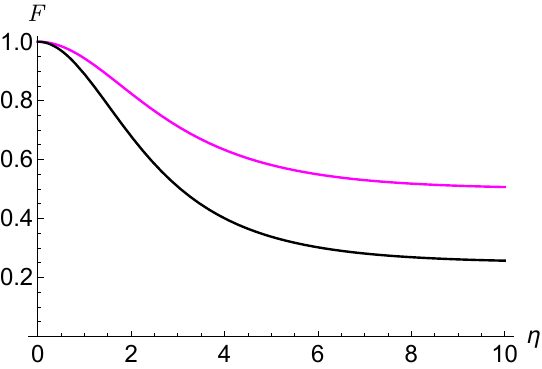}
        \end{minipage}
        \caption{\small Fidelity between the $q$-deformed states and their corresponding undeformed states as a function
of the deformation parameter.  Left: Cases involving $\ket{V_4^{-1}}_{q}$ (salmon), $\ket{V_4^{0}}_{q}$ (purple), $\ket{V_4^{1}}_{q}$ (gray) and $\ket{T_4^{0}}_{q}$ (darker green). Right: Cases involving $\ket{R_4^{\pm 1}}_{q}$ (pink) and $\ket{R_4^{0}}_{q}$ (black). Notice that  $F(|R^{0}_{4}\rangle_{q=1},|R^{0}_{4}\rangle_{q})    =F(|Q^{0}_{4}\rangle_{q=1},|Q^{0}_{4}\rangle_{q})$.}
        \label{fig:fiq42}
    \end{figure}

    A comparison between the $q$-deformed $4-$qubit states and their ideal undeformed counterparts yields the following fidelities:
    \begin{align}
        F(|D^{-1}_{4}\rangle_{q=1},|D^{-1}_{4}\rangle_{q}) & =F(|D^{1}_{4}\rangle_{q=1},|D^{1}_{4}\rangle_{q})=\frac{\left(\sqrt{q}+1\right)^{2}(q+1)}{4 \left(q^{2}+1\right)}, \notag                             \\
        F(|D^{0}_{4}\rangle_{q=1},|D^{0}_{4}\rangle_{q})   & =F(|M^{0}_{4}\rangle_{q=1},|M^{0}_{4}\rangle_{q})=\frac{(q+1)^{2}\left(q+\sqrt{q}+1\right)}{6 \left(q-\sqrt{q}+1\right) \left(q^{2}+1\right)}, \notag \\
        F(|M^{-1}_{4}\rangle_{q=1},|M^{-1}_{4}\rangle_{q}) & =\frac{\left(q+\sqrt{q}+1\right) \left(q^{3/2}+3 q-3 \sqrt{q}+3\right)^{2}}{12 \left(q-\sqrt{q}+1\right) \left(q^{3}+q^{2}+q+1\right)}, \notag        \\
        F(|M^{1}_{4}\rangle_{q=1},|M^{1}_{4}\rangle_{q})   & = \frac{\left(q+\sqrt{q}+1\right) \left(3 q^{3/2}-3 q+3 \sqrt{q}+1\right)^{2}}{12 (q+1) \left(q-\sqrt{q}+1\right) \left(q^{2}+1\right)}, \notag       \\
        F(|V^{-1}_{4}\rangle_{q=1},|V^{-1}_{4}\rangle_{q}) & = \frac{\left(q^{3/2}+3 q+2\right)^{2}}{6 (q+1) \left(q^{2}+q+1\right)}, \notag                                                                       \\
        F(|V^{0}_{4}\rangle_{q=1},|V^{0}_{4}\rangle_{q})   & = \frac{\left(5 q^{3/2}+q^{2}+5 \sqrt{q}+1\right)^{2}}{12 (q+1)^{2}\left(q^{2}+q+1\right)}, \notag                                                    \\
        F(|V^{1}_{4}\rangle_{q=1},|V^{1}_{4}\rangle_{q})   & = \frac{\left(2 q^{3/2}+3 \sqrt{q}+1\right)^{2}}{6 (q+1) \left(q^{2}+q+1\right)},
    \end{align}
    \begin{align}
        F(|T^{0}_{4}\rangle_{q=1},|T^{0}_{4}\rangle_{q})   & =\frac{\left(q^{3/2}+2 q^{2}+6 q+\sqrt{q}+2\right)^{2}}{12 (q+1)^{2}\left(q^{2}+q+1\right)}, \notag \\
        F(|R^{-1}_{4}\rangle_{q=1},|R^{-1}_{4}\rangle_{q}) & =F(|R^{1}_{4}\rangle_{q=1},|R^{1}_{4}\rangle_{q})=\frac{(\sqrt{q}+1)^{2}}{2 (q+1)}, \notag          \\
        F(|R^{0}_{4}\rangle_{q=1},|R^{0}_{4}\rangle_{q})   & =F(|Q^{0}_{4}\rangle_{q=1},|Q^{0}_{4}\rangle_{q})=\frac{\left(\sqrt{q}+1\right)^{4}}{4 (q+1)^{2}}.
    \end{align}
    These fidelities are represented as functions of the deformation parameter in
    Figs.~\ref{fig:fiq41} and \ref{fig:fiq42}.

The same general trend in fidelity as it increases is observed as in the cases of $N=2$ and 3 qubits. 
    The robustness of the fidelity also depends on the class of entanglement being
    considered. Particular configurations such as $\ket{D_4^{\pm1}}_{q}$ and
    $\ket{D_4^{0}}_{q}$, and especially $\ket{M_4^{-1}}_{q}$ and $\ket{V_4^{0}}_{q}$,
    exhibit the most pronounced reduction in fidelity as $\eta$ increases.

Crucially, these different rates of fidelity decay cannot be explained
solely in terms of the number of computational-basis components appearing in
each state. Rather, the relevant point is how the deformation redistributes the
weight among those components, and in particular whether the component selected
in the strong-deformation regime was already relevant in the undeformed state.

There are two distinct mechanisms that can lead to a slower decay of the
fidelity. The first occurs for states involving a smaller number of
computational-basis components, such as the states $\ket{R_4^{\pm1}}_{q}$. Since
these states are supported on fewer configurations, the selection of one
dominant component by the deformation still leaves a relatively large overlap
with the undeformed state. In this case, robustness is mainly a consequence
of the reduced number of components among which the initial probability weight
is distributed.

The second mechanism is different. A state may contain several components, but
the component favored by the deformation may already carry a large weight in
the undeformed state. This is the case of \( |M_4^{1}\rangle_q \). Although this
state contains four computational-basis components, the deformation drives the
state towards the component that already has the largest amplitude at
\(q=1\). As a result, the fidelity remains large even in the strong-deformation
regime. Therefore, the robustness of \( |M_4^{1}\rangle_q \) does not come from
having fewer components, but from the fact that the deformation selects the
component that was already dominant in the undeformed state.
    
\begin{figure}[H]
        \centering
        \begin{minipage}[t]{0.48\linewidth}
            \centering
            \includegraphics[width=\linewidth]{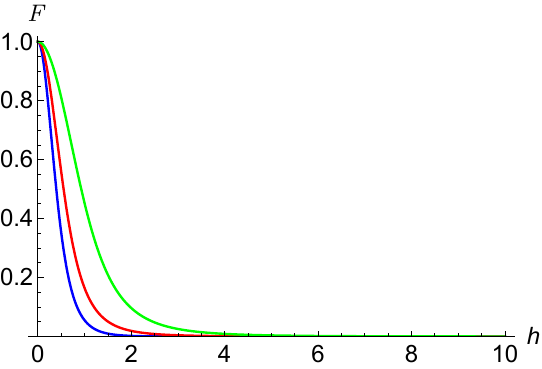}
        \end{minipage}
        \hfill
        \begin{minipage}[t]{0.48\linewidth}
            \centering
            \includegraphics[width=\linewidth]{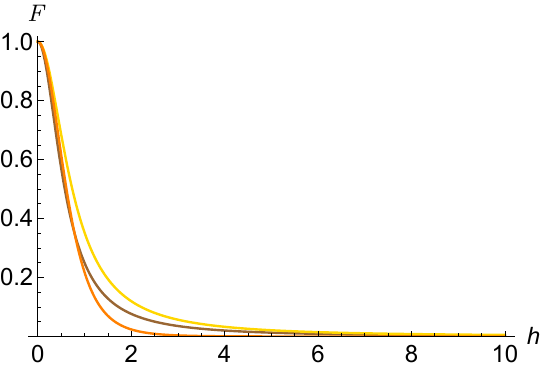}
        \end{minipage}
        \caption{\small Fidelity between the $h$-deformed states and their corresponding undeformed states as a function of the deformation parameter. Left:  Cases involving $\ket{D_4^{-4}}_{h}$ (blue), $\ket{D_4^{-2}}_{h}$ (red) and $\ket{D_4^{0}}_{h}$ (green).  Right:  Cases involving  $\ket{M_4^{-2}}_{h}$ (orange), $\ket{M_4^{0}}_{h}$ (yellow) and $\ket{M_4^{2}}_{h}$ (brown). }
        \label{fig:fih41}
    \end{figure}
\begin{figure}[H]
        \centering
        \begin{minipage}[t]{0.48\linewidth}
            \centering
            \includegraphics[width=\linewidth]{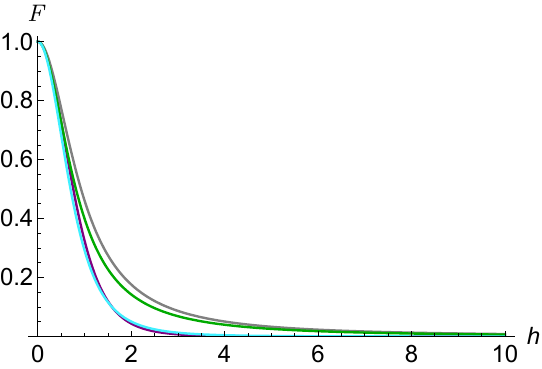}
        \end{minipage}
        \hfill
        \begin{minipage}[t]{0.48\linewidth}
            \centering
            \includegraphics[width=\linewidth]{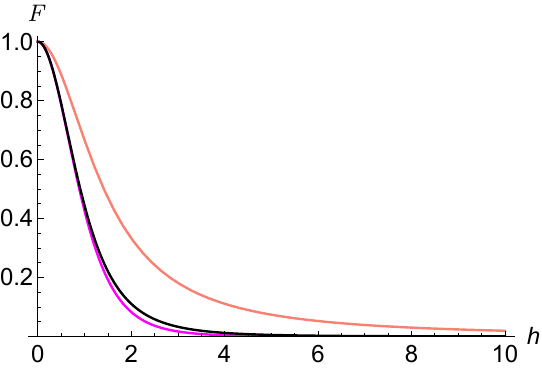}
        \end{minipage}
        \caption{\small Fidelity between the $h$-deformed states and their corresponding undeformed states as a function of the deformation parameter. Left:  Cases involving $\ket{V_4^{-2}}_{h}$ (purple), $\ket{V_4^{0}}_{h}$ (gray), $\ket{V_4^{2}}_{h}$ (darker green) and $\ket{T_4^{0}}_{h}$ (light blue). Right: Cases involving $\ket{R_4^{-2}}_{h}$ (pink), $\ket{R_4^{0}}_{h}$ (salmon) and $\ket{Q_4^{0}}_{h}$ (black). Notice that $ F(|R^{0}_{4}\rangle_{h=0},|R^{0}_{4}\rangle_{h})   =F(|R^{2}_{4}\rangle_{h=0},|R^{2}_{4}\rangle_{h})$. }
        \label{fig:fih42}
    \end{figure}

    On the other hand, fidelities arising when comparing the $h$-deformed states with $4$ qubits with their
    corresponding undeformed counterparts can be written as
    \begin{align}
        F(|D^{-4}_{4}\rangle_{h=0},|D^{-4}_{4}\rangle_{h}) & =\frac{256}{81 h^{8}+848 h^{6}+2144 h^{4}+1280 h^{2}+256}, \notag \\
        F(|D^{-2}_{4}\rangle_{h=0},|D^{-2}_{4}\rangle_{h}) & =\frac{16}{41 h^{4}+40 h^{2}+16}, \notag                          \\
        F(|D^{0}_{4}\rangle_{h=0},|D^{0}_{4}\rangle_{h})   & =\frac{24}{9 h^{4}+20 h^{2}+24}, \notag                           \\
        F(|M^{-2}_{4}\rangle_{h=0},|M^{-2}_{4}\rangle_{h}) & =\frac{48}{9 h^{6}+59 h^{4}+104 h^{2}+48}, \notag                 \\
        F(|M^{0}_{4}\rangle_{h=0},|M^{0}_{4}\rangle_{h})   & =\frac{6}{11 h^{2}+6}, \notag                                     \\
        F(|M^{2}_{4}\rangle_{h=0},|M^{2}_{4}\rangle_{h})   & =\frac{1}{3 h^{2}+1},\notag\\
        F(|V^{-2}_{4}\rangle_{h=0},|V^{-2}_{4}\rangle_{h}) & =\frac{96}{9 h^{6}+62 h^{4}+128 h^{2}+96}, \notag                           \\
        F(|V^{0}_{4}\rangle_{h=0},|V^{0}_{4}\rangle_{h})   & =\frac{6}{7 h^{2}+6}, \notag                                                \\
        F(|V^{2}_{4}\rangle_{h=0},|V^{2}_{4}\rangle_{h})   & =\frac{2}{3 h^{2}+2}, \notag                                                \\
        F(|T^{0}_{4}\rangle_{h=0},|T^{0}_{4}\rangle_{h})   & =\frac{12}{9 h^{4}+20 h^{2}+12}, \notag                                     \\
        F(|R^{-2}_{4}\rangle_{h=0},|R^{-2}_{4}\rangle_{h}) & =\frac{32}{h^{6}+10 h^{4}+32 h^{2}+32}, \notag                              \\
        F(|R^{0}_{4}\rangle_{h=0},|R^{0}_{4}\rangle_{h})   & =F(|R^{2}_{4}\rangle_{h=0},|R^{2}_{4}\rangle_{h})=\frac{2}{h^{2}+2}, \notag \\
        F(|Q^{0}_{4}\rangle_{h=0},|Q^{0}_{4}\rangle_{h})   & =\frac{4}{h^{4}+4 h^{2}+4}.
    \end{align}
    These fidelities are represented as functions of the deformation parameter in
     Figs.~\ref{fig:fih41} and \ref{fig:fih42}. Again, the fidelity decreases
    monotonically as $h$ increases from the undeformed limit ($F=1$) to zero. 
    As the deformation increases, the overlap with the reference state gradually
    decreases. 



As in the 3-qubit case, the rate of fidelity decay strongly depends on the internal structure of the state. Dicke states, such as $\ket{D^{-4}_{4}}_{h}$ and $\ket{D^{-2}_{4}}_{h}$, exhibit a markedly faster decay than other multipartite configurations like $\ket{V^{0}_{4}}_{h}$ and $\ket{R^{2}_{4}}_{h}$. This difference can be traced back to the highly symmetric nature of Dicke states, which are constructed as balanced superpositions of all permutations of the computational-basis configurations with a fixed number of excitations. The Jordanian $h$-deformation, characterized by its asymmetric coproduct and nilpotent contributions, breaks this symmetry and induces a rapid redistribution of amplitudes, thereby suppressing the overlap with the undeformed state. In contrast, states with less symmetric or more heterogeneous structures are less sensitive to this mechanism, as their correlations are not critically tied to a fine-tuned permutation symmetry.
   
    \subsection{Fidelities of \texorpdfstring{$q$}{q}-Dicke states for any \texorpdfstring{$N$}{N} }
While previous sections focused on small system sizes, the algebraic structure of Dicke states allows for a deeper analytical treatment. Unlike non-Dicke states, which require the step-by-step application of the coproduct (or through $q$-Clebsch--Gordan coefficients) without a general closed expression, Dicke states can be constructed via explicit closed-form formulas. This enables us to derive exact analytical expressions for the fidelity under both deformation frameworks that hold for any $N$, including the asymptotic limit $N \to \infty$.  In this subsection, we will focus on the standard deformation, whereas the non-standard one will be addressed in the following subsection.

Consider a one-dimensional spin chain of length $N$ containing $k$
    excitations. The standard Dicke state in the undeformed limit $q \to 1$ is a totally symmetric
    and uniformly weighted superposition of all possible basis states 
    with exactly $k$ excitations, as defined in Eq. \eqref{eq:dicke_def}. So far, we have labelled the Dicke states for the standard 
deformation as $|D_N^m\rangle_q$, where $m$ is the eigenvalue of 
the $L_z$ operator in the corresponding $j = N/2$ irreducible representation. In this Section, for the sake of simplicity in the notation,  we will parameterize the Dicke states by their number of excitations $k$, instead of by the $L_z$ eigenvalue $m$. Thus, we define $\ket{D_N^{(k)}} \equiv \ket{D_N^{m=k-N/2}}$, where $k$ represents the number of up-spins.

    The $q$-deformed analogue of the aforementioned Dicke state, arising for instance from the
    coproduct action of $U_{q}(\mathfrak{sl}(2, \mathbb{R}))$, introduces a
    parameter-dependent weight for each basis configuration. It generates
    coefficients that depend on a weight function $p(S)$ \cite{zhang2009q}:
    \begin{equation}\label{eq31}
         \vert D^{(k)}_{N}\rangle_{q}= \frac{1}{\sqrt{\binom{N}{k}_{q}}}\sum_{S \in
        \mathcal{B}_{N,k}}q^{\frac{1}{2} p(S)}|S\rangle,
    \end{equation}
    where $\binom{N}{k}_{q}$ is the $q$-binomial coefficient defined as
    \begin{align}
        \binom{N}{k}_{q}=\frac{[N]_{q}!}{[N-k]_{q}!\,\, [k]_{q}!},\qquad \textrm{with} \,\,\, [N]_q! = [N]_q [N-1]_q \cdots [1]_q,
    \end{align}
    and $p(S)$ is a configuration-dependent integer exponent that characterizes the
    deformation
    \begin{align}
        p(S)=-k\left( \frac{N+1}{2}\right) + \sum_{i=1}^{k}n_{i},
    \end{align}
    with $n_{i}$ being the position of the $i$-th excitation in the chain.

    Since the $q$-deformed Dicke state is normalized (${}_{q}\langle D^{(k)}_{N}| D^{(k)}_{N}
    \rangle_{q}= 1$) and the basis $\{\vert S \rangle\}$ is orthonormal, the
    following identity strictly holds:
    \begin{equation}
          \sum_{S \in \mathcal{B}_{N,k}}\left( q^{\frac{1}{2} p(S)}\right)^{2}=   \sum_{S \in \mathcal{B}_{N,k}}q^{p(S)}= \binom{N}{k}_{q}.
    \end{equation}
    This establishes a fundamental polynomial identity: the sum of the variable
    raised to the weight function over all permutations generates the deformed
    binomial coefficient in that same variable. This means that for any scalar parameter
    $A$, we have $\sum_{S}A^{p(S)}= \binom{N}{k}_{A}$.

    Now, we evaluate the overlap (probability amplitude)  between the
    standard undeformed state and the $q$-deformed state using their inner
    product:
    \begin{equation}
        \label{eq:overlap_exact} _{q=1}\langle D^{(k)}_{N}\vert D^{(k)}_{N}\rangle
        _{q}= \frac{1}{\sqrt{\binom{N}{k}}}\frac{1}{\sqrt{\binom{N}{k}_{q}}}  \sum_{S \in \mathcal{B}_{N,k}}q^{\frac{1}{2} p(S)}.
    \end{equation}
    Using the polynomial identity from above and evaluating it at the variable $A
    = q^{1/2}$, the summation reduces exactly to:
    \begin{equation}
     \sum_{S \in \mathcal{B}_{N,k}}(q^{1/2})^{p(S)}= \binom{N}{k}_{q^{1/2}}.
    \end{equation}
    Therefore, the overlap becomes:
    \begin{equation}
     _{q=0} \langle D^{(k)}_{N}\vert D^{(k)}_{N}\rangle
        _{q}= \frac{\binom{N}{k}_{q^{1/2}}}{\sqrt{\binom{N}{k}\binom{N}{k}_{q}}}
        .
    \end{equation}
    The fidelity $F$, defined for pure states (as Dicke ones) as the squared modulus of the overlap
    $F = |_{q=0}\langle D^{(k)}_{N}\vert D^{(k)}_{N}\rangle
        _{q}|^{2}$, can thus be expressed
    in a compact form explicitly independent of the exact basis configuration as:
    \begin{equation}
        F(|D^{(k)}_{N}\rangle_{q=1},|D^{(k)}_{N}\rangle_{q}) = \frac{\left[ \binom{N}{k}_{q^{1/2}}\right]^{2}}{\binom{N}{k}\binom{N}{k}_{q}},
        \label{eq:fidelity_q_finite}
    \end{equation}
assuming\footnote{The deformation parameter has been expressed as $q = e^\eta$. When $\eta$ ranges from $0$ to $\infty$, the parameter $q$ naturally maps onto the interval $(1, \infty)$, where $q \to 1^+$ recovers the undeformed limit. Mathematically, however, the algebraic structure of $\mathcal{U}_q(\mathfrak{sl}(2, \mathbb{R}))$ remains perfectly well-defined for any $q \in \mathbb{R}^+ \setminus \{1\}$. If the quantum fidelity is evaluated in the region $0 < q < 1$, the resulting numerical curves exhibit mirror-like behavior or equivalent physical traits to those obtained for $q > 1$. This equivalence is a direct consequence of the $q \leftrightarrow q^{-1}$ invariance inherent in the coproduct of the standard deformation. Therefore, restricting $q$ to the positive real line provides a secure domain to analyze the system's full numerical behavior while avoiding the complex roots of unity, where probability amplitudes would no longer remain real and the quantum fidelity would lose its direct physical interpretation.} $q \in \mathbb{R}^{+}$.  While this formalism gives the exact evaluation for any finite $q$, its
    underlying mechanics dictate the system's behavior in the extreme
    deformation limit, $q \to \infty$.

    In the computational basis representation \eqref{eq31}, the amplitudes scale as
    $q^{\frac{1}{2}p(S)}$. When approaching the $q \to \infty$ threshold, the
    maximally-weighted state is the one maximizing the function $p(S)$. The maximum
    possible value of $p(S)$ for a system of size $N$ with $k$ excitations is exactly
    $k(N-k)/2$. This upper bound is reached by a unique basis state in which all
    the excitations are maximally clustered at one end of the chain, $\vert \down
    \dots\down\up\dots\up \rangle$.

    Because the amplitude of this unique maximal-weight state $|S_{\text{max}}\rangle$
    grows exponentially faster than all other states, the normalization
    naturally drives all competing amplitudes to zero. Physically, this induces
    a spatial ``collapse'' of the $q$-Dicke state. While for $q=1$ the system exists
    as a completely delocalized and highly entangled superposition, at
    $q \to \infty$ it collapses into a completely localized separable product
    state:
    \begin{equation}
        \lim_{q \to \infty}\vert D^{(k)}_{N}\rangle_{q}= \vert S_{\text{max}}\rangle
        .
    \end{equation}

    We can independently verify the analytical limit of the fidelity via our
    exact closed formula, by taking the asymptotic expansion of the $q$-binomial
    for $q \gg 1$:
    \begin{equation}
        \binom{N}{k}_{q}\sim q^{\frac{k(N-k)}{2}}.
    \end{equation}
    This exponent can be rigorously derived by taking the limit of the symmetric
    $q$-number for $q \gg 1$, which behaves as
    \begin{equation}
        [y]_{q}= \frac{q^{y/2}- q^{-y/2}}{q^{1/2}- q^{-1/2}}\sim q^{\frac{y-1}{2}}
        .
    \end{equation}
    Consequently, the asymptotic scaling of the $q$-factorial is given by
    \begin{equation}
        [n]_{q}! \sim \prod_{j=1}^{n}q^{\frac{j-1}{2}}= q^{\frac{n(n-1)}{4}}.
    \end{equation}
    Substituting this expression back into the definition of the $q$-binomial
    coefficient yields
      \begin{equation}
      \binom{N}{k}_{q}= \frac{[N]_{q}!}{[N-k]_{q}! [k]_{q}!}\sim q^{E},     \end{equation}
      where the exponent $E$ is precisely
    \begin{equation}
       E  = \frac{N(N-1)}{4}- \frac{(N-k)(N-k-1)}{4}- \frac{k(k-1)}{4}   = \frac{k(N-k)}{2}.
      \end{equation}

    Applying this limiting behavior to the exact analytical expression derived
    previously:
    \begin{equation}
        \lim_{q \to \infty} F(|D^{(k)}_{N}\rangle_{q=0},|D^{(k)}_{N}\rangle_{q})  = \frac{\left[ (q^{1/2})^{\frac{k(N-k)}{2}}\right]^{2}}{\binom{N}{k}\,\, q^{\frac{k(N-k)}{2}}}
        = \frac{q^{\frac{k(N-k)}{2}}}{\binom{N}{
k
        }\,\, q^{\frac{k(N-k)}{2}}}= \frac{1}{\binom{N}{k}}
        .
        \label{eq:fidelity_limit_q_inf}
    \end{equation}
    The $q$-dependent factors cancel out, leading to a purely
    combinatorial result.

    From a logical consistency viewpoint, this analytical result perfectly aligns
    with the physical interpretation of the state collapse (Kashiwara crystal limit). If the deformed
    state in the $q \to \infty$ limit approaches a single fully localized basis
    state $\vert S_{\text{max}}\rangle$, and the non-deformed reference state ($q
    =1$) is a uniform maximal superposition over all $\binom{N}{k}$ possible configurations,
    the transition overlap is simply the geometric projection of the uniform
    state onto the fully localized state. The squared magnitude of this trivial projection
    yields $1/\binom{N}{k}$ directly. These results explain the numerical
    findings for small $N$.

\subsubsection{Fidelity per site in the thermodynamic limit}

To analyze the thermodynamic limit of the fidelity between the $q$-deformed Dicke state and its undeformed counterpart, we define the corresponding scaling variables. Let $N$ be the system size and $k$ the number of excitations. We introduce the macroscopic excitation density $x = k/N$, where $x \in [0,1]$, and parametrize the quantum deformation as $q = e^\eta$. 

The fidelity per site, $f$, is defined through the thermodynamic rate function of the total fidelity $F$ as\cite{Zhou_2011,Fidelitypersite}
\begin{equation}
    \ln  f(x,\eta) = \lim_{N \to \infty} \frac{1}{N} \ln   F(x,\eta).
\end{equation}
We study this quantity in two distinct scaling regimes of the deformation parameter $\eta$.

\paragraph{Weak deformation regime ($\eta \ll 1$)}

The thermodynamic limit $N\to\infty$ and the weak deformation limit $\eta\to 0$ do not commute, as we will see below from the small-$\eta$ expansion of the exact fidelity of Eq.~\eqref{eq:fidelity_q_finite}. Making an expansion in $\eta$, one finds
\begin{equation}
    [n]_q
    \approx
    n\left[
    1+\frac{\eta^2}{24}(n^2-1)
    +\mathcal O(\eta^4)
    \right],
\end{equation}
so
\begin{equation}
    [n]_{q^{1/2}}
  \approx
    n\left[
    1+\frac{\eta^2}{96}(n^2-1)
    +\mathcal O(\eta^4)
    \right].
\end{equation}
Expanding the logarithm of the symmetric $q$-numbers for $\eta\ll1$, one finds:
\begin{equation}
    \ln [n]_q \approx \ln n + \frac{\eta^2}{24}(n^2 - 1) + \mathcal{O}(\eta^4), \quad \ln [n]_{q^{1/2}} \approx \ln n + \frac{\eta^2}{96}(n^2 - 1) + \mathcal{O}(\eta^4).
\end{equation}
Transforming the operator product of the quantum binomial coefficient $\binom{N}{k}_q$ into a sum of logarithms yields:
\begin{equation}
    \ln \binom{N}{k}_q = \ln \binom{N}{k} + \frac{\eta^2}{24} \left[ \sum_{n=1}^N (n^2 - 1) - \sum_{n=1}^k (n^2 - 1) - \sum_{n=1}^{N-k} (n^2 - 1) \right].
\end{equation}
Evaluating the sums using Faulhaber's formula 
\begin{equation}
S(y) =  \sum_{n=1}^y (n^2-1) = \frac{y(y-1)(2y+5)}{6},
\end{equation} the difference algebraically collapses to $S(N) - S(k) - S(N-k) = k(N - k)(N + 1)$. Introducing the macroscopic filling fraction $x = k/N$, the deformed coefficients are approximated as:
\begin{align}
    \ln \binom{N}{k}_q &= \ln \binom{N}{k} + \frac{\eta^2}{24}N^2(N + 1)x(1 - x) + \mathcal{O}(\eta^4), \\
    \ln \binom{N}{k}_{q^{1/2}} &= \ln \binom{N}{k} + \frac{\eta^2}{96}N^2(N + 1)x(1 - x) + \mathcal{O}(\eta^4).
\end{align}
Therefore, the logarithm of the fidelity behaves as
\begin{equation}
    \frac{1}{N}\ln  F(x,\eta) 
    =
    -\frac{\eta^2}{48}N(N+1)x(1-x)
    +\mathcal O(\eta^4).
\end{equation}

This expression shows that keeping $\eta$ fixed while taking $N\to\infty$ leads to an immediate orthogonality catastrophe, since the intensive logarithmic fidelity diverges as $N^2\eta^2$. Hence, in order to obtain a finite thermodynamic rate function, the deformation parameter must scale inversely with the system size. We therefore introduce the double scaling limit
\begin{equation}
    \eta=\frac{\tilde \eta}{N},
\end{equation}
where $\tilde \eta$ is a finite intensive control parameter. This scaling in the deformation parameter $\eta$ was also derived in \cite{Ballesteros:2025cia} in order to keep the extensibility of the Kittel--Shore model (maintaining a finite quotient between energy and volume in the thermodynamic limit). 

Substituting this scaling into the previous expression gives
\begin{equation}
    \frac{1}{N}\ln F (x,\tilde \eta)
    =
    -\frac{\tilde \eta^2}{48}
    \left(1+\frac{1}{N}\right)
    x(1-x)
    +\mathcal O\!\left(\frac{\tilde \eta^4}{N^2}\right).
\end{equation}
Consequently, in the thermodynamic limit,
\begin{equation}
    \ln f(x,\tilde \eta)
    \equiv
    \lim_{N\to\infty}\frac{1}{N}\ln F (x,\tilde \eta)
    =
    -\frac{\tilde \eta^2}{48}x(1-x)+ \mathcal O\!\left(\tilde \eta^4\right).
\end{equation}

This result identifies the weak-deformation regime as a Gaussian decay regime for the fidelity per site. The decay rate is maximal at half filling, $x=1/2$, where
\begin{equation}
    \ln f\left(\frac{1}{2},\tilde \eta\right)
    =
    -\frac{\tilde \eta^2}{192}.
\end{equation}
Thus the Dicke state with $k=N/2$ excitations presents the lowest fidelity  per site in the weak $q$-deformation regime. This state is also the one which exhibits maximal permutation entropy.  Notice that a Dicke state $\lvert D_N^{(k)} \rangle$ is a uniform superposition (with identical coefficients) of all computational basis states that contain exactly $k$ up-spins (or excitations) and $N-k$ down-spins. The number of states satisfying this condition is given by the binomial coefficient $\binom{N}{k}$. One can associate with an ``entropy'' that measures how dispersed or distributed the state is across the computational basis. At the point $x=1/2$, the Dicke state is composed of the largest possible number of computational basis vectors. By possessing the maximum number of entangled and interchangeable components (permutations), it exhibits maximal permutation entropy.

\paragraph{Strong deformation regime ($\eta\to \infty$)}

In the regime where $\eta$ is held fixed as a large constant while  $N \to \infty$ (which formally includes the extreme limit $\eta \to \infty$), the asymptotic behavior is thus entirely dictated by the asymptotic expansion of the $q$-binomial coefficients in Eq. \eqref{eq:fidelity_q_finite}. Remarkably, for any finite $\eta$ independent of $N$, the leading order terms induced by the quantum deformation exactly cancel out between the numerator and the denominator in Eq. \eqref{eq:fidelity_q_finite}. Consequently, the asymptotic behavior is given by the classical binomial coefficient present in the denominator. Introducing the density $x = k/N$ and applying Stirling's approximation to the binomial coefficient, the logarithm of the total fidelity reads:
\begin{equation}
    \ln F(x) \simeq -\ln \binom{N}{xN} \simeq N \left[ x \ln x + (1-x) \ln(1-x) \right].
\end{equation}
Dividing by the system size $N$, we find that the fidelity per site $\ln f$ collapses to a constant independent of the perturbation strength $\eta$ and entirely determined by the classical Shannon entropy of the system $\mathcal{H}(x)$:
\begin{equation}
    \ln f(x) = x \ln x + (1-x) \ln(1-x) = - \mathcal{H}(x).
\end{equation}
Then, the fidelity decreases as the filling
fraction moves away from the extremal values $x=0$ and $x=1$, since
$\mathcal{H}(0)=\mathcal{H}(1)=0$ and $f=1$. The minimum value is reached at half filling,
$x=1/2$  (i.e. $k=N/2$), where $\mathcal{H}(1/2)=\ln 2$ and therefore
\begin{equation}
\ln f(1/2)=-\ln 2,
\qquad
F(1/2)\sim 2^{-N}.
\end{equation}
Thus, for any fixed filling fraction $0<x<1$, the total fidelity approaches zero in the thermodynamic limit, while the fidelity per site remains finite. Consequently, the weak- and strong-deformation regimes are proven to be different: for small deformations, the fidelity shows a Gaussian dependence on the deformation parameter, whereas the strong-deformation regime is controlled by the binary Shannon entropy $\mathcal{H}(x)$, independently of the value of the deformation parameter.


    \subsection{Fidelities for \texorpdfstring{$h$}{h}-Dicke states for any \texorpdfstring{$N$}{N}}

Quantum fidelities must always be evaluated using properly normalized states to ensure a well-defined physical interpretation. Since the $h$-deformed Clebsch--Gordan coefficients do not, in general,
produce normalized states, we distinguish between the algebraic
(non-normalized) states $\ket{\tilde{\Phi}_h}$ and their normalized counterparts $ \ket{\Phi}_h$, which are related by means of:  
\begin{equation}
    \ket{\Phi}_h
    =
    \frac{\ket{\tilde{\Phi}}_h}
   {\sqrt{_h\langle \tilde{\Phi}| \tilde{\Phi}\rangle_{h}}}.
\end{equation}

The non-normalized  $h$-deformed Dicke states $|\tilde{D}^{-N}_{N}\rangle_{h}$ are expressed as a superposition of the computational basis states $\ket{S}= \ket{s_1 s_2 \dots s_N}$ belonging to the basis $\mathcal{B}_N$ defined in Eq.~\eqref{eq22} as explained in \cite{ballesteros2025entangled}, i.e.:
    \begin{equation}
        \label{eq58}|\tilde D^{-N}_{N}\rangle_{h}= \sum_{S\in \mathcal{B}_N}
        \xi_{n_\uparrow}(P_{\uparrow}(S)) \left(\frac{h}{2}\right)^{n_\uparrow(S)}
        |S\rangle,
    \end{equation}
    where $P_{\uparrow}(S) = \{j \mid s_{j}= \up\}$ is the set of indices
    indicating the positions of the up-spins in $\ket{S}$, and $n_{\uparrow}(S) = |P_{\uparrow}
    (S)|$ is the total number of up-spins in $S$. The structural factor ${\xi}(P_{\uparrow}(S))$
    can be elegantly expressed in terms of elementary symmetric polynomials. Given
    the positional function $g(j) = 2j - N - 1$, we let $e_{m}(P_{\up})$ be the
    $m$-th elementary symmetric polynomial of the values
    $\{g(j) \mid j \in P_{\up}\}$:
    \begin{align}
        e_{0}(P_{\up}) & = 1,\nonumber                                                             \\
        e_{1}(P_{\up}) & = \sum_{j \in P_\up}g(j), \nonumber                                       \\
        e_{2}(P_{\up}) & = \sum_{j_1 < j_2 \in P_\up}g(j_{1})g(j_{2}), \nonumber                   \\
                       & \dots \nonumber                                                           \\
        e_{m}(P_{\up}) & = \sum_{j_1 < j_2 < \dots < j_m \in P_\up}g(j_{1})g(j_{2})\dots g(j_{m}).
    \end{align}
    By convention, $e_{m}(P_{\up}) = 0$ if $m > n_{\up}$. The coefficient $\xi$ is
    completely defined as:
    \begin{equation}
        {\xi}_{n_\up}(P_{\up}) = \sum_{m=0}^{\lfloor n_\up/2 \rfloor}\mathcal C_{m}(N) e_{n_\up-2m}
        (P_{\up}), \label{eq:gamma_general}
    \end{equation}
    where $\mathcal C_{m}(N)$ are particular polynomial coefficients in $N$ ($\mathcal C_{0}(N) = 1$,
    $\mathcal C_{1}(N) = N$, $\mathcal C_{2}(N)=N(3N-2)$), whose leading terms precisely scale as $(
    2m-1)!! N^{m}$ \cite{ballesteros2025entangled}.

    In the undeformed limit ($h \to 0$), the only surviving term is $ |\downarrow\downarrow\dots\downarrow\rangle$, which corresponds to
    zero excitations ($n_{\uparrow}= 0$). The term with zero excitations in $|\tilde{ D}^{-N}_{N}\rangle_{h}$ is the one with coefficient $\xi_{0}= \mathcal C_{0}e_{0}= 1$. Consequently, there is only one nonvanishing contribution to the fidelity, leading to
      \begin{equation}
        \label{eq:fidelity_D}F(|D^{-N}_{N}\rangle_{h=0},|D^{-N}_{N}\rangle_{h})= \frac{|_{h=0}\langle  D^{-N}_{N}| \tilde  D^{-N}_{N}\rangle_{h}|^{2}}{|_h\langle  \tilde D^{-N}_{N}| \tilde D^{-N}_{N}\rangle_{h} |}
        = \frac{1}{\displaystyle \sum_{S \in \mathcal{B}_N}\left[ \xi_{n_\uparrow}(P_{\uparrow}(S)) \right]^{2}\left(\frac{h}{2}\right)^{2n_\uparrow(S)}}
        .
    \end{equation}

   Another interesting state with an analytical closed formula \cite{ballesteros2025entangled} is the $h$-Dicke state for a single excitation, also called $W$-state ($| D^{(1)}_{N}\rangle_{h} \equiv | W_{N}\rangle_{h}$). Its non-normalized form is given by:
    \begin{equation}
        \label{eq64}|\tilde  W_{N}\rangle_{h}= \sum_{S\in \mathcal{B}_N}{\xi}^{\prime}_{n_\uparrow}(P_{\uparrow}
        (S))\left(\frac{h}{2}\right)^{n_\up(S)-1}|S\rangle,
    \end{equation}
    with:
    \begin{align}
        \label{eq65}{\xi}^{\prime}_{n_\uparrow}(P_{\uparrow}) & = \sum_{m=0}^{\lfloor (n_\uparrow-1)/2 \rfloor}(2m+1) \mathcal C_{m}(N) e_{n_\uparrow-1-2m}(P_{\uparrow}) \nonumber  \\
                                                              & - 2 \sum_{m=0}^{\lfloor (n_\uparrow-3)/2 \rfloor}\binom{2m+3}{3}\mathcal C_{m}(N) e_{n_\uparrow-3-2m}(P_{\uparrow}).
    \end{align}
In the limit $h\to 0$, the only surviving contributions of $|\tilde  W_{N}\rangle_{h}$ are those with
$n_\uparrow=1$. Hence,
$\ket{ W_N}_{h=0}$ is the uniform superposition of
the $N$ single-excitation computational basis states:
\begin{equation}
    \ket{W_N}_{h=0}
    =
    \frac{1}{\sqrt{N}}
   \sum_{\substack{S \in \mathcal{B}_N \\ n_\uparrow(S)=1}} \ket S.
\end{equation}
Since there are $N$ terms with one excitation in the deformed state $\ket{\tilde W_N}_h$, the fidelity is
   \begin{equation}
        \label{eq:fidelity_W}F(|W_{N}\rangle_{h=0},|W_{N}\rangle_{h})= \frac{|_{h=0}\langle  W_{N}| \tilde W_{N}\rangle_{h}|^{2}}{|_h\langle  \tilde W_{N}| \tilde W_{N}\rangle_{h} |}
        = \frac{N}{\displaystyle \sum_{S \in \mathcal{B}_N}
\left[\xi'_{n_\uparrow}(P_{\uparrow}(S))\right]^2
\left(\frac{h}{2}\right)^{2n_\uparrow(S)-2}}
        .
    \end{equation}

    It is possible to obtain the rest of $h$-deformed Dicke states for any number of excitations by applying the coproduct operator $\Delta_{h}^{(N)}(Z_{+})$ (Eq.~\ref{eq:coproduct_j}) iteratively starting
    from the deformed ground state $ |\tilde D^{-N}_{N}\rangle_{h}$.  The action of the coproduct is highly non-trivial, but when applied to spin-1/2 representation states $\ket \Psi$, since $Z_{+}^{2}\ket{\down}=0$, it reduces to \cite{ballesteros2025inf}:
    \begin{align}
        \Delta_{h}^{(N)}(Z_{+})\ket{\Psi} & = \sum_{l=1}^{N}\left( 1^{\otimes (l-1)}\otimes Z_{+}\otimes 1^{\otimes (N-l)}\right)\ket{\Psi}\nonumber                                                                                                                              \\
                                          & \quad - \frac{h^{2}}{2}\sum_{1 \le l < j <t \le N}\left( 1^{\otimes (l-1)}\otimes Z_{+}\otimes 1^{\otimes (j-l-1)}\otimes Z_{+}\right.\nonumber \\
                                          & \quad \quad \left.\otimes 1^{\otimes (t-j-1)}\otimes Z_{+}\otimes 1^{\otimes (N-t)}\right)\ket{\Psi}. \label{eq:z+simp}
    \end{align}

This explicit action becomes particularly transparent in the macroscopic
limit ($N\to\infty$), where the structure of the deformed ground state
simplifies. In this regime, the states $|\tilde D^{-N}_{N}\rangle_h$ can be
approximated as \cite{ballesteros2025inf}
    \begin{equation}
        |\tilde D^{-N}_{N}\rangle_{h}= \sum_{S\in \mathcal{B}_N}{\xi}_{n_\uparrow}(S) \left(\frac{h}{2}\right
        )^{n_\uparrow(S)}|S\rangle\approx \sum_{S \in \mathcal{B}_N}\sum_{m=0}^{\lfloor n_\up/2 \rfloor}
        (2m-1)!! N^{m}e_{n_\up-2m}(P_{\up})\left(\frac{h}{2}\right)^{n_\uparrow(S)}
        |S\rangle.
        \label{dicke_fundamental_N}
    \end{equation}
    Then, a clear separation of scales emerges in the coproduct expansion. While
    the structural coefficients scale as $ \mathcal{O}(N^{m})$, the
    elementary symmetric polynomials scale as $e_{n_\up-2m}\sim \mathcal{O}(N^{n_\up-2m})$.
    So the overall scaling behaves as $\mathcal{O}(N^{m}) \times \mathcal{O}
    (N^{n_\up-2m}) = \mathcal{O}(N^{n_\up-m})$. This strict monotonic decrease in the power of $N$ with respect to $m$
implies that the $m=0$ contribution dominates the macroscopic structure.
Thus, at leading order, the dominant state is governed by
$e_{n_\uparrow}(P_\uparrow)$. In the same large-$N$ regime, the term involving the $h$-parameter
in the coproduct \eqref{eq:z+simp} produces contributions involving elementary symmetric
polynomials of lower degree and is therefore subleading with respect to the
undeformed term of $\Delta_{h}^{(N)}(Z_{+})$. This means that if $N \to \infty$, the coproduct \eqref{eq:z+simp} can be well approximated by its undeformed version (with $h=0$). When applying this coproduct operator to the state \eqref{dicke_fundamental_N}, the operation consists of sequentially traversing all positions and adding an up-spin.    This leads to the following closed-form asymptotic expression for the macroscopic Dicke state, strictly valid if the number of times $i$ the coproduct is applied is such that $i \ll N$:
\begin{equation}
\label{eq:asymptotic_dicke_general}
| D_N^{-N+2i}\rangle_h
\propto
\sum_{S\in \mathcal{B}_N}
e_{n_\uparrow(S)-i}(P_\uparrow(S))
\left(\frac{h}{2}\right)^{n_\uparrow(S)-i}
|S\rangle .
\end{equation}

    To systematically quantify the impact of the quantum deformation on the macroscopic
    properties of the system, we derive below the general expression for the fidelity
    $F(_{h=0}\langle D_{N}^{-N+2i}| D_{N}^{-N+2i}\rangle_{h})$. We operate in the asymptotic regime ($N \to \infty $), where the $h$-Dicke
    states can be approximated by Eq. \eqref{eq:asymptotic_dicke_general}.

\subsubsection{Fidelity per site in the thermodynamic limit}


Within this leading thermodynamic approximation, the finite-size fidelity is defined as the squared overlap between the normalized deformed state and
the undeformed Dicke state. Since the latter is a uniform symmetric
superposition of $\binom{N}{i}$ computational basis configurations, and the
deformed state assigns unit weight to each of these configurations in the
limit $h\to 0$, the overlap is entirely determined by their number. Therefore,
one obtains
\begin{equation}
    F( |D_{N}^{-N+2i}\rangle_{h=0}, \,\big|\, D_{N}^{-N+2i}\rangle_{h})
    =
    \frac{\binom{N}{i}}{\mathcal Z_i(h,N)},
\end{equation}
where $\mathcal Z_i(h,N)$  plays the role
of a structural partition function and comes from the squared norm of the corresponding non-normalized deformed state
\begin{equation} 
    \label{eq:partition_function}
    \mathcal{Z}_{i}(h, N)
    =
   \sum_{\substack{S\in\mathcal B_N}}
    \left( \frac{h}{2} \right)^{2(n_\uparrow(S)-i)}
\left[e_{n_\uparrow-i}(P_\uparrow(S))\right]^2.
\end{equation}


 Taking the logarithm and dividing by the system size, one finds the following fidelity per site: 
\begin{equation} 
    \ln f(_{h=0}\langle D_{N}^{-N+2i}| D_{N}^{-N+2i}\rangle_{h}) = \lim_{N \to \infty} \frac{1}{N} \ln F(_{h=0}\langle D_{N}^{-N+2i}| D_{N}^{-N+2i}\rangle_{h})= \lim_{N \to \infty} \left( - \frac{1}{N} \ln \mathcal{Z}_i(h, N) \right), 
\end{equation} 
where we have used the fact that the initial configurational entropy $\frac{1}{N} \ln \binom{N}{i}$ vanishes in the macroscopic scaling $N\to \infty$ for any finite number of initial excitations $i \ll N$. 
For fixed $i$ (or, more generally, for any number of excitations such that
$i/N\to 0$ as $N\to\infty$), the dependence of $\mathcal Z_i(h,N)$ on $i$
only contributes through terms that are negligible after division by $N$.
Therefore, if $i \ll N$ the fidelity per site $f(_{h=0}\langle D_{N}^{-N+2i}| D_{N}^{-N+2i}\rangle_{h})$ coincides with $f(_{h=0}\langle D_{N}^{-N}| D_{N}^{-N}\rangle_{h})$. It is thus sufficient to evaluate the leading contribution coming from $i=0$.

For $i=0$, the relevant term of $\mathcal{Z}_i(h, N)$ is $[e_{n_\up}(P_\uparrow)]^2$. Since $P_\uparrow$ is a
subset of exactly ${n_\up}$ sites, the ${n_\up}$-th elementary symmetric polynomial reduces
to the product of all values in the subset,
\begin{equation}
e_{n_\up}(P_\uparrow)=\prod_{j\in P_\uparrow} g(j),
\end{equation}
so the corresponding partition function becomes
\begin{equation}
    \mathcal Z_0(h,N)
    =
    \sum_{S\in\mathcal B_N}
    \left(\frac{h}{2}\right)^{2n_\uparrow(S)}
    \prod_{j\in P_\uparrow(S)} g(j)^2 .
\end{equation}
By the generating function of elementary symmetric polynomials, this sum
factorizes as
\begin{equation}
    \mathcal Z_0(h,N)
    =
    \prod_{j=1}^{N}
    \left[
    1+\left(\frac{h}{2}\right)^2 g(j)^2
    \right].
\end{equation}
This factorization shows that, at leading order, each lattice site contributes
through a local weight that depends explicitly on its position.

Using $g(j)=2j-N-1$, the fidelity per site is therefore determined by
\begin{equation}
    \ln f(h)
    =
    -\lim_{N\to\infty}
    \frac{1}{N}
    \sum_{j=1}^{N}
    \ln\!\left[
    1+\frac{h^2}{4}(2j-N-1)^2
    \right].
\end{equation}
Introducing the continuous variable $x=j/N$, one has
$2j-N-1\approx N(2x-1)$ in the thermodynamic limit, and the sum becomes
\begin{equation}
    \ln f(h)
   \approx 
    -\int_0^1
    \ln\!\left[
    1+\frac{h^2N^2}{4}(2x-1)^2
    \right]dx .
\end{equation}

This exact integral immediately reveals the conditions for the orthogonality catastrophe. If the deformation $h$ remains a finite constant as $N \to \infty$, the argument of the logarithm grows as $\mathcal{O}(N^2)$, causing the integral to diverge logarithmically and forcing $f(h) \to 0$ (i.e., $\ln f(h) \to -\infty$). The deformed macroscopic state becomes strictly orthogonal to the undeformed state, independently of the fixed value of $h$.  To prevent this catastrophic loss of fidelity and preserve a macroscopic overlap, the deformation parameter must be dynamically rescaled with the system size. The exact mathematical form of the integral precisely dictates the critical scaling: 
\begin{equation} 
    h = \frac{\tilde{h}}{N}, 
\end{equation} 
where $\tilde{h}$ is a system-size independent constant (analogous to the double scaling limit $\eta = \tilde \eta/N$ discussed for the standard deformation). Under this critical scaling, the macroscopic $N$ dependence strictly cancels out. By substituting $y = x - 1/2$, the integral is solved analytically without any truncations or approximations.   Thus, the continuum limit of the leading thermodynamic contribution for the thermodynamic rate function of any finitely-excited initial state ($i \ll N$) is given by the formula: 
\begin{equation} 
  \ln f(\tilde{h}) \approx - 2 \int_{0}^{1/2} \ln \left[ 1 + \tilde{h}^2 y^2 \right] dy = 2 - \ln\left(1 + \frac{\tilde{h}^2}{4}\right) - \frac{4}{\tilde{h}} \arctan\left(\frac{\tilde{h}}{2}\right). 
  \label{eq:exact_fidelity_closed}
\end{equation} 
 We  now analyze the closed-form expression for the fidelity per site of Eq.~\eqref{eq:exact_fidelity_closed} in the two extremal regimes of the rescaled deformation parameter. 

\paragraph{Weak Deformation Regime ($\tilde{h} \ll 1$):} 
performing a Taylor expansion of the logarithmic and arctangent functions around $\tilde{h} = 0$  yields the leading-order behavior of the thermodynamic rate function: 
\begin{equation} 
    \label{eq:weak_h_limit} 
    \ln f(\tilde{h}) \approx 2 - \left(\frac{\tilde{h}^2}{4}\right) - \frac{4}{\tilde{h}} \left( \frac{\tilde{h}}{2} - \frac{\tilde{h}^3}{24} \right) + \mathcal O\!\left(\tilde h^4\right)= - \frac{\tilde{h}^2}{12}+ \mathcal O\!\left(\tilde h^4\right), 
\end{equation} 
showing a Gaussian dependence on the rescaled deformation parameter.
The absence of a linear term means that the first deviation from the
undeformed case is quadratic in $\tilde h$. Therefore, for small
deformations, the fidelity per site remains close to one and decreases
smoothly as $\tilde h$ increases.

\paragraph{Strong Deformation Regime ($\tilde{h} \gg 1$):} 
Conversely, when the rescaled deformation is extremely large, $\arctan(\tilde{h}/2) \to \pi/2$ and $\ln(1 + \tilde{h}^2/4) \approx 2 \ln(\tilde{h}) - \ln(4)$. The thermodynamic rate function simplifies asymptotically to: 
\begin{equation} 
    \label{eq:strong_h_limit} 
    \ln f(\tilde{h}) \approx 2 - 2 \ln(\tilde{h}) + \ln(4) - \frac{2\pi}{\tilde{h}}, 
\end{equation} 
which implies that, for $\tilde h\gg1$,
\begin{equation} 
f(\tilde h)\approx\frac{4e^2}{\tilde h^2}.
\end{equation} 
Thus, in the strong-deformation regime, the fidelity per site decays
algebraically with the rescaled deformation parameter, in contrast with
the Gaussian behaviour found for weak deformation.

    \subsection{Comparison of Thermodynamic Scaling in \texorpdfstring{$q$}{q}- and \texorpdfstring{$h$}{h}-Deformations}
From the fidelity viewpoint, both the standard $q$-deformation and the
Jordanian $h$-deformation require a critical rescaling of the deformation
parameter, of order $N^{-1}$, in order to obtain a finite and non-trivial
fidelity per site in the thermodynamic limit. However, only in the standard
$q$-deformation this rescaling is also required for spectral reasons.

    In the standard Drinfeld-Jimbo $q$-deformation, such as in the $q$-deformed
    Kittel--Shore model \cite{Ballesteros:2025cia}, the deformation modifies the macroscopic
    eigenvalues themselves. Setting $q = e^{\eta}$, the modified spectrum transforms
    proportionally to $\sinh(\eta X/2)/\sinh(\eta/2)$. To prevent the energy
    from diverging exponentially---which would violate thermodynamic extensivity,
    where energy must scale proportionally to the system volume $N$---the argument
    of the hyperbolic sine must remain finite. Consequently, the deformation parameter
    naturally follows the standard intensive mean-field scaling law:
    $\eta =\tilde  \eta/N$. Under this explicit spectral constraint, the system
    remains thermodynamically well-behaved \cite{Ballesteros:2025cia}. The same scaling has been found here by imposing that the fidelity per site goes to a finite (non-vanishing) value. 

    By contrast, the Jordanian $h$-deformation does not change the eigenvalues of the $h$-Kittel--Shore model but it acts purely on the internal combinatorics of the spin lattice. The algebraic operators generate position-dependent factors that break permutation symmetry. As demonstrated by the asymptotic fidelity per site, in the weak-deformation
regime one finds
\[
\ln f(\tilde h)=-\frac{\tilde h^2}{12},
\]
so that the total fidelity scales as
\[
F\sim \exp\left(-\frac{N\tilde h^2}{12}\right).
\]
The role of the critical scaling is not to keep the total fidelity finite
in the thermodynamic limit, but rather to obtain a finite and non-trivial
fidelity per site. Without the scaling $h=\tilde h/N$, the rate function
itself diverges and the fidelity per site collapses. Crucially, this scaling behavior cannot be inferred from the spectrum of the $h$-Kittel--Shore model, as the $h$-deformation leaves the macroscopic energy eigenvalues unmodified.

    \section{Conclusions}
    \label{sec:conclusions}

    In this work, we study the impact of standard ($q$-) and non-standard ($h$-, Jordanian) deformations
    of $\mathcal{U}(\mathfrak{sl}(2, \mathbb{R}))$ on quantum systems of different number of qubits.  We present a comparison between the experimental data and the theoretical predictions for both deformations. When unexpected states are created with negligible probability, the $q$-deformation provides an accurate description by differently reweighting the basis components according to the value of $q$. Conversely, when the probability of these new states becomes non-negligible, the $h$-deformation is required for proper modeling. In particular, for the latter scenario, a combination of both deformations may yield a superior description, as it simultaneously allows for the reweighting of the expected components and the emergence of new ones. A natural candidate is the hybrid $(q,h)$-deformation, obtained by
combining both $q$- and $h$-deformations. Despite this mixed origin, the
hybrid deformation is still of the standard type, since it can be derived
from the standard $q$-deformation through a nonlinear change of basis, either
at the level of the quantum group coordinates~\cite{Aghamohammadi:1994kk} or
of the quantum algebra generators~\cite{Ballesteros_1999}. This possibility
could lead to a more accurate description of the experimental states and will
be explored in future work.

To quantify the impact of these algebraic deformations, we compute the theoretical fidelity between the deformed states and their undeformed counterparts ($q=1$ or $h=0$). This quantum overlap serves as a direct measure of how the state space deviates from the original scenario as the deformation parameters and the number of qubits $N$ vary. In all cases studied, the fidelity decreases with increasing deformation, though displaying distinct behaviors. The $q$-deformation causes a smooth,
    monotonic decay that stabilizes to a non-zero constant, leading to configurations that still retain
    a residual algebraic connection to the initial symmetric structures.
    Conversely, the $h$-deformation induces a faster and stronger loss of fidelity,
    rapidly orthogonalizing the states with respect to the original ones and drastically altering the overall
    correlation architecture. Furthermore, these behavioral differences become
    profound in the macroscopic limit for arbitrary $N$-qubit Dicke states. As $N$
    grows, the $q$-deformation geometrically drives the symmetric superposition
    towards a fully localized, separable product state in the strong-deformation
    regime ($q \to \infty$). Assuring a stable macroscopic limit ($N \to \infty$) thus requires a standard intensive deformation parameter scaling of $N^{-1}$, which agrees with a previous result when imposing the thermodynamic extensivity of the $q$-Kittel--Shore spectrum. Similarly, the $h$-deformation mathematically alters the spatial combinatorics of the spin lattice, inducing an orthogonality catastrophe that also demands a critical scaling of $N^{-1}$ to stabilize the macroscopic quantum states. In this case, this result cannot be obtained from the  thermodynamic extensivity of the $h$-Kittel--Shore spectrum, since its eigenvalues are $h$ independent.

    Moreover, in future articles, we plan to extend this analysis by investigating
    in detail the behavior of various entanglement measures, such as von Neumann
    and Rényi entropies, tangles, and logarithmic negativities under these
    deformations. This will allow for a more complete characterization of the generation
    and suppression of quantum correlations.

    \section*{Acknowledgments}
    The authors acknowledge partial support from the grants PID2023-148373NB-I00 funded
    by MCIN/ AEI / 10.13039/501100011033/FEDER -- UE, as well as BU011P25 funded by the Regional Government of Castilla y Le\'on (Junta de Castilla y Le\'on). V. Mariscal acknowledges support from the Universidad de Burgos through
    its PhD grant program.

    \appendix
    \section{\texorpdfstring{$q$}{q}-deformed states of up to 4 qubits}
    \label{app1:sec1} 
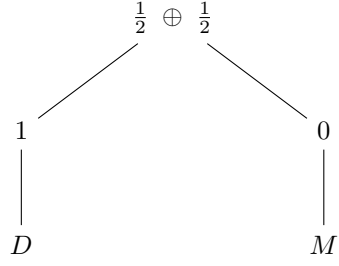
\begin{figure}[H]
        \centering
        \begin{tikzpicture}[
            level distance=1.5cm,
            level 1/.style={sibling distance=4cm},
            every node/.style={font=\small, align=center},
            vertical child/.style={ edge from parent path={(\tikzparentnode.south) -- (\tikzchildnode.north)} }
        ]
            \node {$\tfrac{1}{2}\;\oplus\;\tfrac{1}{2}$}
                child
                { node {$1$} child[vertical child] { node {$D$} } }
                child
                { node {$0$} child[vertical child] { node {$M$} } };
        \end{tikzpicture}
        \caption{Coupling tree for two spin-$\tfrac12$ particles, showing $J=1,0$
        states, labeled as $D$ and $M$, respectively~\cite{ballesteros2025entangled}.}
        \label{fig:two_spin_tree_DU}
    \end{figure}

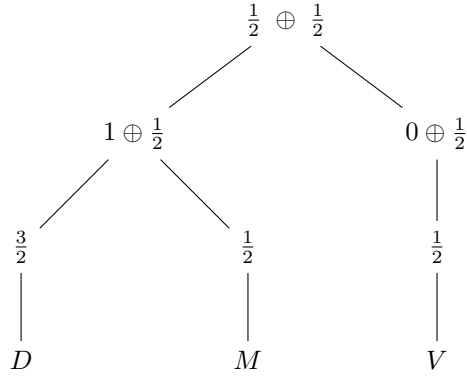
\begin{figure}[H]
        \centering
        \begin{tikzpicture}[
            level distance=1.5cm,
            level 1/.style={sibling distance=4cm},
            level 2/.style={sibling distance=3cm},
            every node/.style={font=\small, align=center},
            vertical child/.style={ edge from parent path={(\tikzparentnode.south) -- (\tikzchildnode.north)} }
        ]
            \node {$\tfrac{1}{2}\;\oplus\;\tfrac{1}{2}$}
                child
                { node {$1 \oplus\tfrac{1}{2}$}
                child { node {$\tfrac{3}{2}$} child[vertical child] { node {$D$} } } child { node {$\tfrac{1}{2}$} child[vertical child] { node {$M$} } } }
                child
                { node {$0\oplus \tfrac{1}{2}$}
                child { node {$\tfrac{1}{2}$} child[vertical child] { node {$V$} } } };
        \end{tikzpicture}
        \caption{Coupling tree for three spin-$\tfrac{1}{2}$ particles, with labels
        $D$, $M$ and $V$~\cite{ballesteros2025entangled}. }
        \label{fig:three_spin_tree_DUV}
    \end{figure}

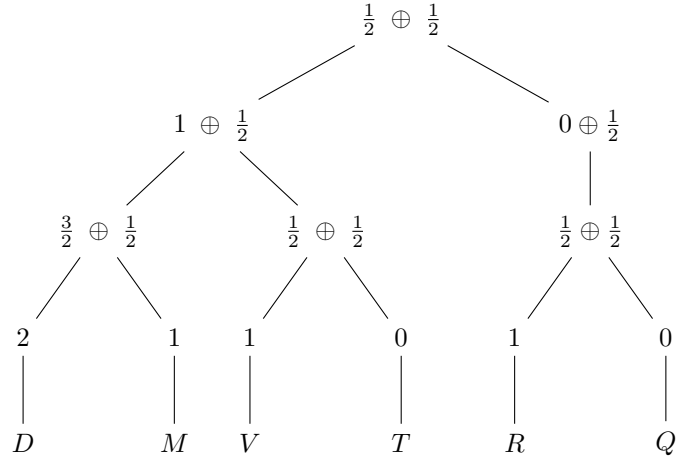
\begin{figure}[h!]
        \centering
        \begin{tikzpicture}[
            level distance=1.4cm, 
            level 1/.style={sibling distance=5cm}, 
            level 2/.style={sibling distance=3cm}, 
            level 3/.style={sibling distance=2cm}, 
            every node/.style={font=\small, align=center}, 
            vertical child/.style={ edge from parent path={(\tikzparentnode.south) -- (\tikzchildnode.north)} }
        ]
            \node {$\tfrac{1}{2}\;\oplus\;\tfrac{1}{2}$}
                child
                { node {$1\;\oplus\;\tfrac{1}{2}$}
                child { node {$\tfrac{3}{2}\;\oplus\;\tfrac{1}{2}$}
                child { node {$2$}
                child[vertical child] { node {$D$} } } child { node {$1$} child[vertical child] { node {$M$} } } }
                child { node {$\tfrac{1}{2}\;\oplus\;\tfrac{1}{2}$}
                child { node {$1$} child[vertical child] { node {$V$} } } child { node {$0$} child[vertical child] { node {$T$} } } } }
                child
                { node {$0 \oplus \tfrac{1}{2}$}
                child { node {$\tfrac{1}{2}\oplus \tfrac{1}{2}$}
                child { node {$1$} child[vertical child] { node {$R$} } } child { node {$0$} child[vertical child] { node {$Q$} } } }};
        \end{tikzpicture}
        \caption{Coupling tree for four spin-$\tfrac{1}{2}$ particles, with labels
        $D$, $M$, $V$, $T$, $R$ and $Q$~\cite{ballesteros2025entangled}.}
        \label{fig:four_spin_tree}
    \end{figure}

  In this section, we collect the normalized states associated with the irreducible representations of the standard 
deformation $\mathcal{U}_q \left(\mathfrak{sl}(2, \mathbb{R})\right)$. We will use the notation 
 $|X_N^m\rangle_q$, where $N$ is the number of qubits (which corresponds mathematically to the $N$-fold tensor product of the fundamental spin-$1/2$ representation), and $m$ the eigenvalue of the $L_z$ operator in the corresponding $j = N/2$ irreducible representation. We also use the computational basis \begin{eqnarray}
&& \mathcal{B}_N=
\left\{
\ket{s_1s_2\cdots s_N}\;:\;
s_\ell\in\{\uparrow,\downarrow\},\ \ell=1,\ldots,N  \right\},\nonumber\\
&&\textrm{where}\quad \ket{s_1s_2\cdots s_N}
=
\ket{s_1}\otimes\ket{s_2}\otimes\cdots\otimes\ket{s_N}.
\end{eqnarray}
This basis constitutes an orthonormal one of the composite Hilbert space $\mathcal{H}_N = (\mathbb{C}^2)^{\otimes N}$. We also define 
\begin{equation}
\left\{ |\uparrow\rangle := \begin{pmatrix} 1 \\ 0 \end{pmatrix}, \quad 
|\downarrow\rangle := \begin{pmatrix} 0 \\ 1 \end{pmatrix} \right\}.
\end{equation}
For $2$ qubits, we use the notation present in Fig.
    \ref{fig:two_spin_tree_DU} for labelling the states.
\noindent The triplet of $q$-Dicke states ($j=1$) is given by:
    \begin{eqnarray}
        && \ket{D^{-1}_2}_q = \ket{\down \down}, \nonumber\\
        &&\ket{D^{0}_2}_q = \frac{1}{\sqrt{[2]_{q}}} (q^{1/4}\ket{\down\up} + q^{-1/4}
        \ket{\up \down}), \nonumber\\
        &&\ket{D^1_2}_q =\ket{\up\up}, \label{eq:2qqubits}
    \end{eqnarray}
    while the singlet ($j=0$) can be written as:
    \begin{eqnarray}
        \ket{M^0_2}_q =\frac{1}{\sqrt{[2]_{q}}} \left(-q^{-1/4}\left|\downarrow\uparrow\right\rangle
        + q^{1/4}\left|\uparrow \downarrow\right\rangle\right).
    \end{eqnarray}
For $3$ qubits, we use the notation present in Fig.~\ref{fig:three_spin_tree_DUV}
    to label the states.
    The quadruplet of $q$-Dicke states (with $j=3/2$) is:
    \begin{eqnarray}
        &&\ket{D^{-3/2}_3}_q =|\hspace{-0.1cm}\down\down\down\rangle, \nonumber\\
        &&\ket{D^{-1/2}_3}_q = \frac{1}{\sqrt{[3]_{q}}}\left( q^{1/2}|\hspace{-0.1cm}\down\down\up\rangle
        +|\hspace{-0.1cm}\down\up\down\rangle+ q^{-1/2}|\hspace{-0.1cm}\up\down\down\rangle\right),
        \nonumber\\
        &&\ket{D^{1/2}_3}_q =\frac{1}{\sqrt{[3]_{q}}}\left(q^{1/2} |\hspace{-0.1cm}\down\up\up\rangle
        +|\hspace{-0.1cm}\up\down\up\rangle + q^{-1/2}|\hspace{-0.1cm}\up\up\down\rangle
        \right),\nonumber\\
        &&\ket{D^{3/2}_3}_q =\left|\up\up\up\right\rangle, \label{eq:3qubitsq}
    \end{eqnarray}
    and the two doublets of $q$-states (with $j=1/2$) can be expressed as
    \begin{eqnarray}
        &&|M^{-1/2}_{3}\rangle_{q}=\frac{1}{\sqrt{[2]_{q}}\sqrt{[3]_{q}}}\left(q^{1/4}\left|\uparrow\downarrow\downarrow\right\rangle+q^{3/4}\left|\downarrow\uparrow\downarrow\right\rangle-[2]_{q}\,\, q^{-1/4}\left|\downarrow\downarrow\uparrow\right\rangle\right),
        \nonumber\\
        &&|M^{1/2}_{3}\rangle_{q}=\frac{1}{\sqrt{[2]_{q}}\sqrt{[3]_{q}}}\left([2]_{q}\,\, q^{1/4}\left|\uparrow\uparrow\downarrow\right\rangle-q^{-3/4}\left|\uparrow\downarrow\uparrow\right\rangle-q^{-1/4}\left|\downarrow\uparrow\uparrow\right\rangle\right),
        \nonumber\\
        &&|V^{-1/2}_{3}\rangle_q=\frac{1}{\sqrt{[2]_{q}}}\left(q^{1/4}\left|\uparrow\downarrow\downarrow\right\rangle-q^{-1/4}\left|\downarrow\uparrow\downarrow\right\rangle\right),
        \nonumber\\
        &&|V^{1/2}_{3}\rangle_q=\frac{1}{\sqrt{[2]_{q}}}\left(q^{1/4}\left|\uparrow\downarrow\uparrow\right\rangle-q^{-1/4}\left|\downarrow\uparrow\uparrow\right\rangle\right).
    \end{eqnarray}
    The corresponding deformed states for a $4$-qubit system are presented here for the first time, labeled according to Fig. \ref{fig:four_spin_tree}.  
Using the $q$-Clebsch--Gordan coefficients~\eqref{eq:qcg}, we obtain the
    quintuplet of $q$-Dicke states (with $j=2$)
    \begin{eqnarray}
        &&|D^{-2}_{4}\rangle_{q}=\left|\downarrow\downarrow\downarrow\downarrow\right\rangle,
        \nonumber\\
        &&|D_{4}^{-1}\rangle_{q}=\frac{1}{\sqrt{[4]_{q}}}\left(q^{-3/4}\left|\uparrow\downarrow\downarrow\downarrow\right\rangle+q^{-1/4}\left|\downarrow\uparrow\downarrow\downarrow\right\rangle+q^{1/4}\left|\downarrow\downarrow\uparrow\downarrow\right\rangle+q^{3/4}\left|\downarrow\downarrow\downarrow\uparrow\right\rangle\right),
        \nonumber\\
        &&|D^{0}_{4}\rangle_{q}=\frac{\sqrt{[2]_{q}}}{\sqrt{[3]_{q}}\sqrt{[4]_{q}}}\left(q^{-1}\left|\uparrow\uparrow\downarrow\downarrow\right\rangle+q^{-1/2}\left|\uparrow\downarrow\uparrow\downarrow\right\rangle+\left|\uparrow\downarrow\downarrow\uparrow\right\rangle+\left|\downarrow\uparrow\uparrow\downarrow\right\rangle+q^{1/2}\left|\downarrow\uparrow\downarrow\uparrow\right\rangle+q\left|\downarrow\downarrow\uparrow\uparrow\right\rangle\right),
        \nonumber\\
        &&|D^{1}_{4}\rangle_{q}=\frac{1}{\sqrt{[4]_{q}}}\left(q^{3/4}\left|\downarrow\uparrow\uparrow\uparrow\right\rangle+q^{1/4}\left|\uparrow\downarrow\uparrow\uparrow\right\rangle+q^{-1/4}\left|\uparrow\uparrow\downarrow\uparrow\right\rangle+q^{-3/4}\left|\uparrow\uparrow\uparrow\downarrow\right\rangle\right),
        \nonumber\\
        &&|D_{4}^{2}\rangle_{q}=\left|\uparrow\uparrow\uparrow\uparrow\right\rangle,
    \end{eqnarray}
    together with three triplets (with $j=1$)
    \begin{eqnarray}
        &&|M_{4}^{-1}\rangle_{q}=\frac{1}{\sqrt{[3]_{q}}\sqrt{[4]_{q}}}\left(q^{1/4}\left|\uparrow\downarrow\downarrow\downarrow\right\rangle+q^{3/4}\left|\downarrow\uparrow\downarrow\downarrow\right\rangle+q^{5/4}\left|\downarrow\downarrow\uparrow\downarrow\right\rangle-q^{-1/4}[3]_{q}\left|\downarrow\downarrow\downarrow\uparrow\right\rangle\right),
        \nonumber\\
        &&|M_{4}^{0}\rangle_{q}=\frac{\sqrt{[2]_{q}}}{\sqrt{[3]_{q}}\sqrt{[4]_{q}}}(\left|\uparrow\uparrow\downarrow\downarrow\right\rangle+q^{1/2}\left|\uparrow\downarrow\uparrow\downarrow\right\rangle-q^{-1}\left|\uparrow\downarrow\downarrow\uparrow\right\rangle+q\left|\downarrow\uparrow\uparrow\downarrow\right\rangle-q^{-1/2}\left|\downarrow\uparrow\downarrow\uparrow\right\rangle-\left|\downarrow\downarrow\uparrow\uparrow\right\rangle),
        \nonumber\\
        &&|M_{4}^{1}\rangle_{q}=\frac{1}{\sqrt{[3]_{q}}\sqrt{[4]_{q}}}\left(q^{1/4}[3]_{q}\left|\uparrow\uparrow\uparrow\downarrow\right\rangle-q^{-5/4}\left|\uparrow\uparrow\downarrow\uparrow\right\rangle-q^{-3/4}\left|\uparrow\downarrow\uparrow\uparrow\right\rangle-q^{-1/4}\left|\downarrow\uparrow\uparrow\uparrow\right\rangle\right),\nonumber\\
        &&|V_{4}^{-1}\rangle_{q}=\frac{1}{\sqrt{[2]_{q}}\sqrt{[3]_{q}}}\left(q^{1/4}\left|\uparrow\downarrow\downarrow\downarrow\right\rangle+q^{3/4}\left|\downarrow\uparrow\downarrow\downarrow\right\rangle-q^{-1/4}[2]_{q}\left|\downarrow\downarrow\uparrow\downarrow\right\rangle\right),
        \nonumber\\
        &&|V_{4}^{0}\rangle_{q}=\frac{1}{[2]_{q}\sqrt{[3]_{q}}}\left([2]_{q}\left|\uparrow\uparrow\downarrow\downarrow\right\rangle-q^{-1}\left|\uparrow\downarrow\uparrow\downarrow\right\rangle+q^{1/2}\left|\uparrow\downarrow\downarrow\uparrow\right\rangle-q^{-1/2}\left|\downarrow\uparrow\uparrow\downarrow\right\rangle+q\left|\downarrow\uparrow\downarrow\uparrow\right\rangle-[2]_{q}\left|\downarrow\downarrow\uparrow\uparrow\right\rangle\right),
        \nonumber\\
        &&|V_{4}^{1}\rangle_{q}=\frac{1}{\sqrt{[2]_{q}}\sqrt{[3]_{q}}}\left(q^{1/4}[2]_{q}\left|\uparrow\uparrow\downarrow\uparrow\right\rangle-q^{-3/4}\left|\uparrow\downarrow\uparrow\uparrow\right\rangle-q^{-1/4}\left|\downarrow\uparrow\uparrow\uparrow\right\rangle\right),\nonumber\\
        &&|R_{4}^{-1}\rangle_{q}=\frac{1}{\sqrt{[2]_{q}}}\left(q^{1/4}\left|\uparrow\downarrow\downarrow\downarrow\right\rangle-q^{-1/4}\left|\downarrow\uparrow\downarrow\downarrow\right\rangle\right),
        \nonumber\\
        &&|R_{4}^{0}\rangle_{q}=\frac{1}{[2]_{q}}\left(\left|\uparrow\downarrow\uparrow\downarrow\right\rangle+q^{1/2}\left|\uparrow\downarrow\downarrow\uparrow\right\rangle-q^{-1/2}\left|\downarrow\uparrow\uparrow\downarrow\right\rangle-\left|\downarrow\uparrow\downarrow\uparrow\right\rangle\right),
        \nonumber\\
        &&|R_{4}^{1}\rangle_{q}=\frac{1}{\sqrt{[2]_{q}}}\left(q^{1/4}\left|\uparrow\downarrow\uparrow\uparrow\right\rangle-q^{-1/4}\left|\downarrow\uparrow\uparrow\uparrow\right\rangle\right),
    \end{eqnarray}
    and two singlets (with $j=0$)
    \begin{eqnarray}
        &&|T_{4}^{0}\rangle_{q}=\frac{1}{[2]_{q}\sqrt{[3]_{q}}}\left(q^{1/2}[2]_{q}\left|\uparrow\uparrow\downarrow\downarrow\right\rangle-q^{-1/2}\left|\uparrow\downarrow\uparrow\downarrow\right\rangle-\left|\uparrow\downarrow\downarrow\uparrow\right\rangle-\left|\downarrow\uparrow\uparrow\downarrow\right\rangle-q^{1/2}\left|\downarrow\uparrow\downarrow\uparrow\right\rangle+q^{-1/2}[2]_{q}\left|\downarrow\downarrow\uparrow\uparrow\right\rangle\right),\nonumber\\
        &&|Q_{4}^{0}\rangle_{q}=\frac{1}{[2]_{q}}\left(q^{1/2}\left|\uparrow\downarrow\uparrow\downarrow\right\rangle-\left|\uparrow\downarrow\downarrow\uparrow\right\rangle-\left|\downarrow\uparrow\uparrow\downarrow\right\rangle+q^{-1/2}\left|\downarrow\uparrow\downarrow\uparrow\right\rangle\right).
    \end{eqnarray}

    \section{\texorpdfstring{$h$}{h}-deformed states of up to 4 qubits}
    \label{app1:sec2} 
    
In this section, we collect the normalized states associated with the irreducible representations of the non-standard deformation \(\mathcal{U}_h(\mathfrak{sl}(2,\mathbb{R}))\). We will use the notation \(\ket{X_N^{\tilde m}}_h\), where \(N\) is the number of qubits, corresponding mathematically to the \(N\)-fold tensor product of the fundamental spin-\(1/2\) representation, and \(\tilde m\) is the eigenvalue of the \(H=2J_z\) operator in the corresponding \(j=N/2\) irreducible representation. The computational basis used here is the same as the one introduced in Appendix~\ref{app1:sec1}.
 
\noindent For 2 qubits, the triplet of $h$-Dicke states (with
    $j=1$) is given by:
    \begin{eqnarray}
        &&\ket{D^{-2}_2}_h =\frac{4}{h^{2}+4} \left(\frac{h^{2}}{4}\ket{\up\up}
        -\frac{h}{2}\ket{\up \down}+\frac{h}{2}\ket{ \down\up} + \ket{\down \down}\right),
        \nonumber\\
        &&\ket{D^{0}_2}_h = \frac{1}{\sqrt{2}} (\ket{\up \down}+\ket{ \down\up}),
       \nonumber\\
        &&\ket{D^2_2}_h =\ket{\up\up}, \label{eq:2qubits}
    \end{eqnarray}
    while the singlet state (with $j=0$) is:
    \begin{equation}
        \ket{M^0_2}_{h}=\frac{1}{\sqrt{h^{2}+2}}\left(-h |\up\up\rangle + \ket{\up \down}
        -\ket{ \down\up}\right). \label{Mh}
    \end{equation}
    For $3$ qubits, we obtain the following quadruplet of $h$-Dicke states (with $j=3/2$):
    \begin{eqnarray}
        &&\ket{D^{-3}_3}_h =\frac{4}{\sqrt{19h^{4}+32h^{2}+16}}\left(\frac{3h^{2}}{4}\ket{\up\up\down}
        -\frac{h^{2}}{4}\ket{\up \down\up}-h\ket{ \up\down\down} +\frac{3h^{2}}{4}\ket{\down\up\up}
        +h\ket{ \down\down\up}+ \ket{\down\down \down}\right), \nonumber\\
        &&\ket{D^{-1}_3}_h = \frac{4 \sqrt{3}}{\sqrt{9h^{4}+32h^{2}+48}}\left(\frac{\sqrt{3}h^{2}}{4}\ket{\up\up\up}
        -\frac{h}{\sqrt{3}}\ket{\up\up \down} +\frac{h}{\sqrt{3}} \ket{\down \up\up}+
        \frac{1}{\sqrt{3}} (\ket{ \up\down\down} + \ket{ \down\up \down} +\ket{ \down\down\up}
        )\right), \nonumber\\
        &&\ket{D^1_3}_h =\frac{1}{\sqrt{3}} (\ket{ \up\up\down} + \ket{ \up\down\up }
        +\ket{ \down\up\up} ),\nonumber\\
        &&\ket{D^3_3}_h =|\up\up\up\rangle, \label{eq:3qubitsh}
    \end{eqnarray}
    together with two doublets of $h$-states (with $j=1/2$):
    \begin{eqnarray}
        &&\ket{M^{-1}_3}_h = \frac{\sqrt{6}}{\sqrt{5h^{2}+6}}\left(\frac{1}{\sqrt{6}}
        (-h\ket{ \up\up\down}-2h\ket{ \down\up\up}) +\frac{1}{\sqrt{6}}( \ket{ \up\down\down }
        + \ket{ \down\up\down}-2 \ket{ \down\down\up})\right),\nonumber\\
        &&\ket{M^{1}_3}_h= \frac{\sqrt{2}}{\sqrt{3h^{2}+2}}\left(-\frac{3h}{\sqrt{6}}
        \ket{\up\up \up} +\frac{1}{\sqrt{6}} (2\ket{\up\up \down}-\ket{\up\down \up}-\ket{\down\up \up})\right),\nonumber\\[2.5ex]
        &&\ket{V^{-1}_3}_h=\frac{\sqrt{2}}{\sqrt{h^{2}+2}}\left(-\frac{h}{\sqrt{2}}
        \ket{ \up\up\down}+\frac{1}{\sqrt{2}} ( -\ket{ \down\up\down}+\ket{ \up\down\down})
        \right),\nonumber\\
        &&\ket{V^{1}_3}_h = \frac{\sqrt{2}}{\sqrt{h^{2}+2}}\left(-\frac{h}{\sqrt{2}}
        \ket{ \up\up\up}+\frac{1}{\sqrt{2}} (- \ket{ \down\up\up}+\ket{ \up\down\up})\right).
        \label{eq:3qubitsa}
    \end{eqnarray}
    For $4$ qubits, the quintuplet of $h$-Dicke states (with $j=2$) can be written as:
    \begin{eqnarray}
        &&|D^{-4}_{4}\rangle_{h}=\frac{1}{\sqrt{81 h^{8}+848 h^{6}+2144 h^{4}+1280
        h^{2}+256}}\left(9h^4\left|\uparrow\uparrow\uparrow\uparrow\right\rangle
        -18h^3\left|\uparrow\uparrow\uparrow\downarrow\right\rangle +10h^3\left|\uparrow\uparrow\downarrow\uparrow\right\rangle
        \right. \nonumber\\
        &&\qquad \qquad \left.+28h^2\left|\uparrow\uparrow\downarrow\downarrow\right\rangle -10h^3\left|\uparrow\downarrow\uparrow\uparrow\right\rangle
        +4h^2\left|\uparrow\downarrow\uparrow\downarrow\right\rangle -20h^2\left|\uparrow\downarrow\downarrow\uparrow\right\rangle
        -24h\left|\uparrow\downarrow\downarrow\downarrow\right\rangle +18h^3\left|\downarrow\uparrow\uparrow\uparrow\right\rangle
        \right.\nonumber \\
        &&\qquad \qquad\left.+12h^2\left|\downarrow\uparrow\uparrow\downarrow\right\rangle +4h^2\left|\downarrow\uparrow\downarrow\uparrow\right\rangle
        -8h\left|\downarrow\uparrow\downarrow\downarrow\right\rangle +28h^2\left|\downarrow\downarrow\uparrow\uparrow\right\rangle
        +8h\left|\downarrow\downarrow\uparrow\downarrow\right\rangle +24h\left|\downarrow\downarrow\downarrow\uparrow\right\rangle
        +16\left|\downarrow\downarrow\downarrow\downarrow\right\rangle \right), \nonumber\\
        &&|D_{4}^{-2}\rangle_{h}=\frac{1}{2 \sqrt{41 h^{4}+40 h^{2}+16}}\left(9h^2\left|\uparrow\uparrow\uparrow\downarrow\right\rangle
        +h^2\left|\uparrow\uparrow\downarrow\uparrow\right\rangle -8h\left|\uparrow\uparrow\downarrow\downarrow\right\rangle
        +h^2\left|\uparrow\downarrow\uparrow\uparrow\right\rangle -4h\left|\uparrow\downarrow\uparrow\downarrow\right\rangle
        \right.\nonumber \\
        &&\qquad \qquad\left.+4\left|\uparrow\downarrow\downarrow\downarrow\right\rangle +9h^2\left|\downarrow\uparrow\uparrow\uparrow\right\rangle
        +4h\left|\downarrow\uparrow\downarrow\uparrow\right\rangle +4\left|\downarrow\uparrow\downarrow\downarrow\right\rangle
        +8h\left|\downarrow\downarrow\uparrow\uparrow\right\rangle +4\left|\downarrow\downarrow\uparrow\downarrow\right\rangle
        +4\left|\downarrow\downarrow\downarrow\uparrow\right\rangle\right),
        \nonumber\\
       && |D^{0}_{4}\rangle_{h}=\frac{1}{\sqrt{9 h^{4}+20 h^{2}+24}}\left(3h^2\left|\uparrow\uparrow\uparrow\uparrow\right\rangle
        -3h\left|\uparrow\uparrow\uparrow\downarrow\right\rangle -h\left|\uparrow\uparrow\downarrow\uparrow\right\rangle
        +2\left|\uparrow\uparrow\downarrow\downarrow\right\rangle +h\left|\uparrow\downarrow\uparrow\uparrow\right\rangle
        +2\left|\uparrow\downarrow\uparrow\downarrow\right\rangle\right. \nonumber
        \\
        &&\qquad \qquad \left.+2\left|\uparrow\downarrow\downarrow\uparrow\right\rangle +3h\left|\downarrow\uparrow\uparrow\uparrow\right\rangle
        +2\left|\downarrow\uparrow\uparrow\downarrow\right\rangle +2\left|\downarrow\uparrow\downarrow\uparrow\right\rangle
        +2\left|\downarrow\downarrow\uparrow\uparrow\right\rangle\right),
        \nonumber\\
        &&|D^{2}_{4}\rangle_{h}=\frac{1}{2}|\uparrow\uparrow\uparrow\downarrow\rangle
        +\frac{1}{2}|\uparrow\uparrow\downarrow\uparrow\rangle +\frac{1}{2}|\uparrow\downarrow\uparrow\uparrow\rangle
        +\frac{1}{2}|\downarrow\uparrow\uparrow\uparrow\rangle, \nonumber\\
        &&|D_{4}^{4}\rangle_{h}=|\uparrow\uparrow\uparrow\uparrow\rangle,
    \end{eqnarray}
    while the three triplets (with $j=1$) are
    \begin{eqnarray}
       && |M_{4}^{-2}\rangle_{h}=\frac{1}{2 \sqrt{9 h^{6}+59 h^{4}+104 h^{2}+48}}\left(-6h^3\left|\uparrow\uparrow\uparrow\uparrow\right\rangle
        +9h^2\left|\uparrow\uparrow\uparrow\downarrow\right\rangle -3h^2\left|\uparrow\uparrow\downarrow\uparrow\right\rangle
        -8h\left|\uparrow\uparrow\downarrow\downarrow\right\rangle \right.\nonumber
        \\
        &&\qquad \qquad+5h^2\left|\uparrow\downarrow\uparrow\uparrow\right\rangle -4h\left|\uparrow\downarrow\uparrow\downarrow\right\rangle
        +8h\left|\uparrow\downarrow\downarrow\uparrow\right\rangle +4\left|\uparrow\downarrow\downarrow\downarrow\right\rangle
        -11h^2\left|\downarrow\uparrow\uparrow\uparrow\right\rangle -4h\left|\downarrow\uparrow\downarrow\uparrow\right\rangle
        +4\left|\downarrow\uparrow\downarrow\downarrow\right\rangle\nonumber \\
        &&\qquad \qquad\left.-16h\left|\downarrow\downarrow\uparrow\uparrow\right\rangle +4\left|\downarrow\downarrow\uparrow\downarrow\right\rangle
        -12\left|\downarrow\downarrow\downarrow\uparrow\right\rangle \right),
        \nonumber\\
        &&|M_{4}^{0}\rangle_{h}=\frac{1}{2\sqrt{11 h^{2}+6}} \left(-3h\left|\uparrow\uparrow\uparrow\downarrow\right\rangle
        -h\left|\uparrow\uparrow\downarrow\uparrow\right\rangle +2\left|\uparrow\uparrow\downarrow\downarrow\right\rangle
        -3h\left|\uparrow\downarrow\uparrow\uparrow\right\rangle +2\left|\uparrow\downarrow\uparrow\downarrow\right\rangle
        -2\left|\uparrow\downarrow\downarrow\uparrow\right\rangle\right.
        \nonumber \\
        &&\qquad \qquad\left.-5h\left|\downarrow\uparrow\uparrow\uparrow\right\rangle +2\left|\downarrow\uparrow\uparrow\downarrow\right\rangle
        -2\left|\downarrow\uparrow\downarrow\uparrow\right\rangle -2\left|\downarrow\downarrow\uparrow\uparrow\right\rangle\right),
        \nonumber\\
        &&|M_{4}^{2}\rangle_{h}=-\frac{1}{2 \sqrt{9 h^{2}+3}}\left(6h\left|\uparrow\uparrow\uparrow\uparrow\right\rangle
        -3\left|\uparrow\uparrow\uparrow\downarrow\right\rangle +\left|\uparrow\uparrow\downarrow\uparrow\right\rangle
        +\left|\uparrow\downarrow\uparrow\uparrow\right\rangle +\left|\downarrow\uparrow\uparrow\uparrow\right\rangle\right),\nonumber\\
       && |V_{4}^{-2}\rangle_{h}=\frac{1}{\sqrt{9 h^{6}+62 h^{4}+128 h^{2}+96}}\left(-3h^3\left|\uparrow\uparrow\uparrow\uparrow\right\rangle
        +6h^2\left|\uparrow\uparrow\uparrow\downarrow\right\rangle -8h\left|\uparrow\uparrow\downarrow\downarrow\right\rangle
        -h^2\left|\uparrow\downarrow\uparrow\uparrow\right\rangle +2h\left|\uparrow\downarrow\uparrow\downarrow\right\rangle
        \right. \nonumber \\
        &&\qquad \qquad\left.+2h\left|\uparrow\downarrow\downarrow\uparrow\right\rangle +4\left|\uparrow\downarrow\downarrow\downarrow\right\rangle
        -5h^2\left|\downarrow\uparrow\uparrow\uparrow\right\rangle -6h\left|\downarrow\uparrow\uparrow\downarrow\right\rangle
        +2h\left|\downarrow\uparrow\downarrow\uparrow\right\rangle +4\left|\downarrow\uparrow\downarrow\downarrow\right\rangle
        -4h\left|\downarrow\downarrow\uparrow\uparrow\right\rangle -8\left|\downarrow\downarrow\uparrow\downarrow\right\rangle
        \right), \nonumber\\
       && |V_{4}^{0}\rangle_{h}=-\frac{1}{\sqrt{2} \sqrt{7 h^{2}+6}}\left(3h\left|\uparrow\uparrow\uparrow\downarrow\right\rangle
        +h\left|\uparrow\uparrow\downarrow\uparrow\right\rangle -2\left|\uparrow\uparrow\downarrow\downarrow\right\rangle
        +\left|\uparrow\downarrow\uparrow\downarrow\right\rangle -\left|\uparrow\downarrow\downarrow\uparrow\right\rangle
        +2h\left|\downarrow\uparrow\uparrow\uparrow\right\rangle +\left|\downarrow\uparrow\uparrow\downarrow\right\rangle
        \right. \nonumber \\
        &&\qquad \qquad\left.-\left|\downarrow\uparrow\downarrow\uparrow\right\rangle +2\left|\downarrow\downarrow\uparrow\uparrow\right\rangle\right),
        \nonumber\\
        &&|V_{4}^{2}\rangle_{h}=\frac{1}{\sqrt{9 h^{2}+6}}\left(-3h\left|\uparrow\uparrow\uparrow\uparrow\right\rangle
        +2\left|\uparrow\uparrow\downarrow\uparrow\right\rangle -\left|\uparrow\downarrow\uparrow\uparrow\right\rangle
        -\left|\downarrow\uparrow\uparrow\uparrow\right\rangle\right),\nonumber\\
        &&|R_{4}^{-2}\rangle_{h}=\frac{1}{\left(h^{2}+4\right)\sqrt{h^2+2}}\left(-h^3\left|\uparrow\uparrow\uparrow\uparrow\right\rangle
        +2h^2\left|\uparrow\uparrow\uparrow\downarrow\right\rangle -2h^2\left|\uparrow\uparrow\downarrow\uparrow\right\rangle
        -4h\left|\uparrow\uparrow\downarrow\downarrow\right\rangle +h^2\left|\uparrow\downarrow\uparrow\uparrow\right\rangle
        \right. \nonumber \\
        &&\qquad \qquad\left.-2h\left|\uparrow\downarrow\uparrow\downarrow\right\rangle +2h\left|\uparrow\downarrow\downarrow\uparrow\right\rangle
        +4\left|\uparrow\downarrow\downarrow\downarrow\right\rangle -h^2\left|\downarrow\uparrow\uparrow\uparrow\right\rangle
        +2h\left|\downarrow\uparrow\uparrow\downarrow\right\rangle -2h\left|\downarrow\uparrow\downarrow\uparrow\right\rangle
        -4\left|\downarrow\uparrow\downarrow\downarrow\right\rangle\right),\nonumber\\
        &&|R_{4}^{0}\rangle_{h}=-\frac{1}{\sqrt{2} \sqrt{h^{2}+2}}\left(h\left|\uparrow\uparrow\uparrow\downarrow\right\rangle
        +h\left|\uparrow\uparrow\downarrow\uparrow\right\rangle -\left|\uparrow\downarrow\uparrow\downarrow\right\rangle
        -\left|\uparrow\downarrow\downarrow\uparrow\right\rangle +\left|\downarrow\uparrow\uparrow\downarrow\right\rangle
        +\left|\downarrow\uparrow\downarrow\uparrow\right\rangle\right), \nonumber\\
       && |R_{4}^{2}\rangle_{h}=\frac{1}{\sqrt{h^{2}+2}}\left(-h\left|\uparrow\uparrow\uparrow\uparrow\right\rangle
        +\left|\uparrow\downarrow\uparrow\uparrow\right\rangle -\left|\downarrow\uparrow\uparrow\uparrow\right\rangle\right).
    \end{eqnarray}
    Finally, we also obtain two singlets (with $j=0$):
    \begin{eqnarray}
        &&|T_{4}^{0}\rangle_{h}=\frac{1}{\sqrt{9 h^{4}+20 h^{2}+12}}\left(3h^2\left|\uparrow\uparrow\uparrow\uparrow\right\rangle
        -3h\left|\uparrow\uparrow\uparrow\downarrow\right\rangle -h\left|\uparrow\uparrow\downarrow\uparrow\right\rangle
        +2\left|\uparrow\uparrow\downarrow\downarrow\right\rangle +h\left|\uparrow\downarrow\uparrow\uparrow\right\rangle
        -\left|\uparrow\downarrow\uparrow\downarrow\right\rangle \right.
        \nonumber \\
        &&\qquad \qquad\left.-\left|\uparrow\downarrow\downarrow\uparrow\right\rangle +3h\left|\downarrow\uparrow\uparrow\uparrow\right\rangle
        -\left|\downarrow\uparrow\uparrow\downarrow\right\rangle -\left|\downarrow\uparrow\downarrow\uparrow\right\rangle
        +2\left|\downarrow\downarrow\uparrow\uparrow\right\rangle\right), \nonumber\\
        &&|Q_{4}^{0}\rangle_{h}=\frac{1}{h^{2}+2}\left(h^2\left|\uparrow\uparrow\uparrow\uparrow\right\rangle
        -h\left|\uparrow\uparrow\uparrow\downarrow\right\rangle +h\left|\uparrow\uparrow\downarrow\uparrow\right\rangle
        -h\left|\uparrow\downarrow\uparrow\uparrow\right\rangle +\left|\uparrow\downarrow\uparrow\downarrow\right\rangle
        -\left|\uparrow\downarrow\downarrow\uparrow\right\rangle +h\left|\downarrow\uparrow\uparrow\uparrow\right\rangle
        -\left|\downarrow\uparrow\uparrow\downarrow\right\rangle \right. \nonumber
        \\
        &&\qquad \qquad \left.+\left|\downarrow\uparrow\downarrow\uparrow\right\rangle\right).
    \end{eqnarray}


\begin{thebibliography}{10}

\bibitem{Drinfeld}
V.~G. Drinfel'd.
\newblock Quantum groups.
\newblock In A.~Gleason, editor, {\em Proc. Int. Congr. Math. (Berkeley 1986)},
  pages 798--820, Providence RI, 1987. American Mathematical Society.

\bibitem{Jimbo:1985zk}
M.~Jimbo.
\newblock A {$q$}-difference analog of {$U(g)$} and the {Yang--Baxter}
  equation.
\newblock {\em Lett. Math. Phys.}, 10:63--69, 1985.

\bibitem{gawedzki1991classical}
K.~Gawedzki.
\newblock Classical origin of quantum group symmetries in
  {Wess--Zumino--Witten} conformal field theory.
\newblock {\em Commun. Math. Phys.}, pages 201--213, 1991.

\bibitem{isaev1994quantum}
A.~P. Isaev.
\newblock Quantum group covariant noncommutative geometry.
\newblock {\em J. Math. Phys.}, 35(12):6784--6801, 1994.

\bibitem{drinfel1990hopf}
V.~G. Drinfel'd.
\newblock Hopf algebras and the quantum {Yang--Baxter} equation.
\newblock In Michio Jimbo, editor, {\em {Yang--Baxter} Equation in Integrable
  Systems}, volume~10 of {\em Advanced Series in Mathematical Physics}, pages
  264--268. World Scientific, Singapore, 1990.

\bibitem{kassel2012quantum}
C.~Kassel.
\newblock {\em Quantum Groups}, volume 155.
\newblock Springer New York, New York, NY, 1995.

\bibitem{Chari}
V.~Chari and A.~N. Pressley.
\newblock {\em A guide to quantum groups}.
\newblock Cambridge University Press, Cambridge, 1994.

\bibitem{kulish1993quantum}
P.~P. Kulish.
\newblock Quantum groups, {$q$} oscillators, and covariant algebras.
\newblock {\em Theor. Math. Phys.}, 94(2):137--141, 1993.

\bibitem{kulish1991general}
P.~P. Kulish and E.~K. Sklyanin.
\newblock The general {$\mathcal{U}_{q} (\mathfrak{sl}(2))$} invariant {XXZ}
  integrable quantum spin chain.
\newblock {\em J. Phys. A: Math. Gen.}, 24(8):L435--L439, 1991.

\bibitem{biedenharn1989quantum}
L.~C. Biedenharn.
\newblock The quantum group {$\textit{SU}_{q}$}(2) and a {$q$}-analogue of the
  boson operators.
\newblock {\em J. Phys. A: Math. Gen.}, 22(18):L873--L878, 1989.

\bibitem{kirillov1990uq}
A.~Kirillov and N.~Reshetikhin.
\newblock Representations of the algebra {$\mathcal{U}_{q}
  (\mathfrak{sl}(2))$}, {$q$}-orthogonal polynomials and invariants of links.
\newblock In V.~G. Kac, editor, {\em Infinite Dimensional Lie Algebras and
  Groups}, volume~7 of {\em Advanced Series in Mathematical Physics}, pages
  285--339. World Scientific, Singapore, 1989.

\bibitem{Majid:1988we}
S.~Majid.
\newblock Hopf algebras for physics at the {Planck} scale.
\newblock {\em Class. Quantum Grav.}, 5:1587--1606, 1988.

\bibitem{KulishReshetikhin}
P.~P. Kulish and N.~Yu. Reshetikhin.
\newblock Quantum linear problem for the {Sine--Gordon} equation and higher
  representations.
\newblock {\em J. Soviet Math.}, 23(4):2435--2441, 1983.

\bibitem{AngelBallesteros_1996}
A.~Ballesteros and F.~J. Herranz.
\newblock Universal {$R$}-matrix for non-standard quantum {$sl(2,\mathbb{R})$}.
\newblock {\em J. Phys. A: Math. Gen.}, 29(13):L311, 1996.

\bibitem{ballesteros2025entangled}
A.~Ballesteros, J.~J. Relancio, and L.~Santamaría-Sanz.
\newblock Non-standard quantum algebra {$\mathcal{U}_h (\mathfrak{sl}(2,
  \mathbb{R}))$} and {$h$}-{Dicke} states.
\newblock {\em Quant. Inf. Proc.}, 25(1):12, 2026.

\bibitem{kittel1965development}
C.~Kittel and H.~Shore.
\newblock Development of a phase transition for a rigorously solvable many-body
  system.
\newblock {\em Phys. Rev.}, 138(4A):A1165--A1169, 1965.

\bibitem{Ballesteros:2025cia}
A.~Ballesteros, I.~Gutierrez-Sagredo, V.~Mariscal, and J.~J. Relancio.
\newblock Quantum group deformation of the {Kittel--Shore} model.
\newblock {\em J. Phys. A: Math. Theor.}, 58(44):445202, 2025.

\bibitem{ciftja1999equation}
O.~Ciftja, M.~Luban, M.~Auslender, and J.~H. Luscombe.
\newblock Equation of state and spin-correlation functions of ultrasmall
  classical {Heisenberg} magnets.
\newblock {\em Phys. Rev. B}, 60(14):10122--10133, 1999.

\bibitem{sheng1981melting}
P.~Sheng, R.~W. Cohen, and J.~R. Schrieffer.
\newblock Melting transition of small molecular clusters.
\newblock {\em J. Phys. C: Solid State Phys.}, 14(20):L565--L569, 1981.

\bibitem{ciftja2001irregular}
O.~Ciftja.
\newblock The irregular tetrahedron of classical and quantum spins subjected to
  a magnetic field.
\newblock {\em J. Phys. A: Math. Gen.}, 34(8):1611--1627, 2001.

\bibitem{ciftja2000spin}
O.~Ciftja.
\newblock Spin correlation functions of some frustrated ultra-small classical
  heisenberg clusters.
\newblock {\em Physica A: Stat. Mech. and its Applications}, 286(3-4):541--557,
  2000.

\bibitem{al2004nonlinear}
H.~Al-Wahsh, D.~Bria, A.~Akjouj, and P.~Zieliński.
\newblock Nonlinear effect of perpendicular magnetic field on the
  antiferromagnetic phase transition in weakly coupled layered systems: Equal
  access decoupling scheme.
\newblock {\em Phys. Rev. B}, 70(1):014405, 2004.

\bibitem{al2004extended}
H.~Al-Wahsh, G.~Ismail, and K.~Lotfy.
\newblock Extended random phase approximation for layered copper oxides
  antiferromagnets.
\newblock {\em Czechoslovak J. of Phys.}, 54(12):1511--1520, 2004.

\bibitem{gros1995transition}
C.~Gros, W.~Wenzel, and J.~Richter.
\newblock The transition from an ordered antiferromagnet to a quantum
  disordered spin liquid in a solvable bilayer model.
\newblock {\em Europhys. Lett.}, 32(9):747--752, 1995.

\bibitem{czachor2001green}
A.~Czachor and H.~Al-Wahsh.
\newblock Green’s function approach to the neutron-inelastic-scattering
  determination of magnon dispersion relations for isotropic disordered
  magnets.
\newblock {\em Phys. Rev. B}, 63(6):064419, 2001.

\bibitem{sandoval2015thermodynamic}
S.~M. Sandoval, A.~E. Sepulveda, and S.~M. Keller.
\newblock On the thermodynamic efficiency of a nickel-based multiferroic
  thermomagnetic generator: From bulk to atomic scale.
\newblock {\em J. Appl. Phys.}, 117(16), 2015.

\bibitem{hsu2011thermomagnetic}
C.~Hsu, S.~M. Sandoval, K.~P. Wetzlar, and G.~P. Carman.
\newblock Thermomagnetic conversion efficiencies for ferromagnetic materials.
\newblock {\em J. Appl. Phys.}, 110(12), 2011.

\bibitem{sun2004coordination}
C.~Q. Sun, W.~H. Zhong, S.~Li, B.~K. Tay, H.~L. Bai, and E.~Y. Jiang.
\newblock Coordination imperfection suppressed phase stability of
  ferromagnetic, ferroelectric, and superconductive nanosolids.
\newblock {\em J. Phys. Chem. B}, 108(3):1080--1084, 2004.

\bibitem{ferguson2002ammonia}
A.~J. Ferguson, P.~A. Cain, D.~A. Williams, and G.~A.~D. Briggs.
\newblock Ammonia-based quantum computer.
\newblock {\em Phys. Rev. A}, 65(3):034303, 2002.

\bibitem{woodworth2006few}
R.~Woodworth, A.~Mizel, and D.~A. Lidar.
\newblock Few-body spin couplings and their implications for universal quantum
  computation.
\newblock {\em J. Phys.: Condens. Matter}, 18(21):S721--S744, 2006.

\bibitem{benenti2004principles}
G.~Benenti, G.~Casati, and G.~Strini.
\newblock {\em Principles of Quantum Computation and Information}, volume~1.
\newblock World Scientific Publishing Co. Pte. Ltd., 1 edition, 2004.

\bibitem{barfknecht2019realizing}
R.~E. Barfknecht, S.~E. Rasmussen, A.~Foerster, and N.~T. Zinner.
\newblock Realizing time crystals in discrete quantum few-body systems.
\newblock {\em Phys. Rev. B}, 99(14):144304, 2019.

\bibitem{nielsen_chuang_2010}
M.~A. Nielsen and I.~L. Chuang.
\newblock {\em Quantum Computation and Quantum Information: 10th Anniversary
  Edition}.
\newblock Cambridge University Press, Cambridge, 2010.

\bibitem{jozsa1994}
R.~Jozsa.
\newblock Fidelity for mixed quantum states.
\newblock {\em J. Mod. Opt.}, 41(12):2315--2323, 1994.

\bibitem{dicke1954coherence}
R.~H. Dicke.
\newblock Coherence in spontaneous radiation processes.
\newblock {\em Phys. Rev.}, 93:99--110, 1954.

\bibitem{nepomechie2023qudit}
R.~I. Nepomechie and D.~Raveh.
\newblock Qudit {Dicke} state preparation.
\newblock {\em Quantum Inf. Comput.}, 24(1--2):37--56, 2024.

\bibitem{bartschi2019deterministic}
A.~Bärtschi and S.~Eidenbenz.
\newblock Deterministic preparation of {Dicke} states.
\newblock volume 11651, pages 126--139. 2019.

\bibitem{mukherjee2020actual}
C.~S. Mukherjee, S.~Maitra, V.~Gaurav, and D.~Roy.
\newblock On actual preparation of {Dicke} state on a quantum computer.
\newblock {\em arXiv:2007.01681 [quant-ph]}, 2020.

\bibitem{aktar2022divide}
S.~Aktar, A.~Bärtschi, A.~A. Badawy, and S.~Eidenbenz.
\newblock A divide-and-conquer approach to {Dicke} state preparation.
\newblock {\em IEEE Trans. Quantum Eng.}, 3:1--16, 2022.

\bibitem{Wieczorek2009}
W.~Wieczorek et~al.
\newblock Experimental entanglement of a six-photon symmetric {Dicke} state.
\newblock {\em Phys. Rev. Lett.}, 103(2):020504, 2009.

\bibitem{Prevedel2009}
R.~Prevedel et~al.
\newblock Experimental realization of {Dicke} states of up to six qubits for
  multiparty quantum networking.
\newblock {\em Phys. Rev. Lett.}, 103(2):020503, 2009.

\bibitem{bengtsson2017geometry}
I.~Bengtsson and K.~Zyczkowski.
\newblock {\em Geometry of Quantum States: An Introduction to Quantum
  Entanglement}.
\newblock Cambridge University Press, Cambridge, 2006.

\bibitem{walter2016multipartite}
M.~Walter, D.~Gross, and J.~Eisert.
\newblock Multipartite entanglement.
\newblock {\em Quantum Inf.: From Foundations to Quantum Technology
  Applications}, pages 293--330, 2016.

\bibitem{Sakuraibook}
J.~J. Sakurai and J.~Napolitano.
\newblock {\em Modern Quantum Mechanics}.
\newblock Cambridge University Press, Singapore, 2020.

\bibitem{cohenbook}
C.~Cohen-Tannoudji, B.~Diu, and F.~Lalo{\"e}.
\newblock {\em Quantum Mechanics}.
\newblock Wiley-VCH, 2nd edition, 2020.

\bibitem{Biedenharnbook}
L.~C. Biedenharn and M.~A. Lohe.
\newblock {\em {Quantum Group Symmetry and {$Q$}-Tensor Algebras}}.
\newblock World Scientific Pub Co Inc, Singapore, 1996.

\bibitem{alvarez2024russian}
R.~Álvarez Nodarse and A.~Arenas-Gómez.
\newblock The {$q$}-analog of the quantum theory of angular momentum: a review
  from special functions.
\newblock {\em Russ. J. Math. Phys.}, 31(1):24--43, 2024.

\bibitem{li2015entanglement}
Z.~Li and A.~Wang.
\newblock Entanglement entropy in quasi-symmetric multi-qubit states.
\newblock {\em Int. J. Quantum Inf.}, 13(02):1550007, 2015.

\bibitem{JVanderJeugt1998}
J.~Van der Jeugt.
\newblock Representations and {Clebsch--Gordan} coefficients for the
  {Jordanian} quantum algebra.
\newblock {\em J. Phys. A: Math. Gen.}, 31:1495, 1998.

\bibitem{Aizawa1997}
N.~Aizawa.
\newblock Irreducible decomposition for tensor product representations of
  {Jordanian} quantum algebras.
\newblock {\em J. Phys. A: Math. Gen.}, 30:5981, 1997.

\bibitem{magyari1987integrable}
E.~Magyari, H.~Thomas, R.~Weber, C.~Kaufman, and G.~Muller.
\newblock Integrable and nonintegrable classical spin clusters.
\newblock {\em Z. Phys. B: Condens. Matter.}, 65(3):363--374, 1987.

\bibitem{miller2013classical}
W.~Miller, S.~Post, and P.~Winternitz.
\newblock Classical and quantum superintegrability with applications.
\newblock {\em J. Phys. A: Math. Theor.}, 46(42):423001, 2013.

\bibitem{Faddeev:1996iy}
L.~D. Faddeev.
\newblock How algebraic {Bethe} ansatz works for integrable model.
\newblock In {\em Les Houches School of Physics: Astrophysical Sources of
  Gravitational Radiation}, pages 149--219, 1996.

\bibitem{ballesteros1998systematic}
A.~Ballesteros and O.~Ragnisco.
\newblock A systematic construction of completely integrable {Hamiltonians}
  from coalgebras.
\newblock {\em J. Phys. A: Math. Gen.}, 31(16):3791--3813, 1998.

\bibitem{ballesteros1999integrable}
N.~Ballesteros and F.~J. Herranz.
\newblock Integrable deformations of oscillator chains from quantum algebras.
\newblock {\em J. Phys. A: Math. Gen.}, 32(50):8851--8862, 1999.

\bibitem{ballesteros2009super}
A.~Ballesteros, A.~Blasco, F.~J. Herranz, F.~Musso, and O.~Ragnisco.
\newblock {(Super)integrability} from coalgebra symmetry: Formalism and
  applications.
\newblock {\em J. Phys.: Conf. Ser.}, 175:012004, 2009.

\bibitem{cruz2019efficient}
D.~Cruz et~al.
\newblock Efficient quantum algorithms for {$GHZ$} and {$W$} states, and
  implementation on the {IBM} quantum computer.
\newblock {\em Adv. Quantum Technol.}, 2(5-6):1900015, 2019.

\bibitem{Ballesteros:2025dbv}
A.~Ballesteros, I.~Gutierrez-Sagredo, J.~de~Ramon, and J.~J. Relancio.
\newblock Deformations of the symmetric subspace of qubit chains.
\newblock {\em J. Phys. A}, 58(40):405301, 2025.

\bibitem{kashiwara1991crystal}
M.~Kashiwara.
\newblock On crystal bases of the {$q$}-analogue of universal enveloping
  algebras.
\newblock {\em Duke Math. J.}, 63(2):465--516, 1991.

\bibitem{zhang2009q}
W.~Zhang and G.~Guo.
\newblock A {$q$}-deformation of {Dicke} states.
\newblock {\em Phys. Scr.}, 79(5):055702, 2009.

\bibitem{Zhou_2011}
H.~Q. Zhou, J.~H. Zhao, H.~L. Wang, and B.~Li.
\newblock Singularities in fidelity surfaces for quantum phase transitions: a
  geometric perspective.
\newblock {\em J. Phys. A: Math. Theor.}, 44(4):042002, 2010.

\bibitem{Fidelitypersite}
M.~M. Rams and B.~Damski.
\newblock Quantum fidelity in the thermodynamic limit.
\newblock {\em Phys. Rev. Lett.}, 106(5):055701, 2011.

\bibitem{ballesteros2025inf}
A.~Ballesteros, R.~Ramírez, and M.~Reboiro.
\newblock Non-standard quantum algebras and infinite-dimensional {PT}-symmetric
  systems.
\newblock {\em J. Phys. A}, 58(45):455301, 2025.

\bibitem{Aghamohammadi:1994kk}
A.~Aghamohammadi, M.~Khorrami, and A.~Shariati.
\newblock {$h$}-deformation as a contraction of {$q$}-deformation.
\newblock {\em J. Phys. A}, 28:L225, 1995.

\bibitem{Ballesteros_1999}
A.~Ballesteros, F.~Herranz, and P.~Parashar.
\newblock Multiparametric quantum {$gl(2)$}: Lie bialgebras, quantum
  {$R$}-matrices and non-relativistic limits.
\newblock {\em J. Phys. A: Math. Gen.}, 32(12):2369, 1999.

\end{thebibliography}
\end{document}